\documentclass[11pt,a4paper]{article}
\pdfoutput=1
\usepackage{jcappub}
\usepackage{graphicx}
\usepackage{subfig}
\usepackage{enumerate}

\def\apj{{\it Astrophys.~J.}}

\def\aj{{\it Astronom.~J.}}
\def\apjl{{\it Astrophys.~J.~Lett.}}
\def\prd{{\it Phys.~Rev.~D}}
\def\prl{{\it Phys.~Rev.~Lett.}}
\def\plb{{\it Phys.~Letts.~B}}
\def\mnras{{\it Mon.~Not. Roy.~Astr.~Soc.}}
\def\ijmpd{{\it Int.~J.~ Mod. Phys. D}}

\def\AnA{{\it Astron. Astrophys.}}

\def\ARAnA{{\it Ann. Rev. Astron. Astrophys.}}

\title{Exploring scalar field dynamics with Gaussian processes} 
\author[a]{Remya Nair,}
\author[a]{Sanjay Jhingan}
\author[b]{and Deepak Jain}

\affiliation[a]{Centre for Theoretical Physics,\\ Jamia Millia
Islamia, New Delhi 110025, India}

\affiliation[b]{Deen Dayal Upadhyaya College, \\
University of Delhi, New Delhi 110015, India}

\emailAdd{remya$_{-}$phy@yahoo.com}
\emailAdd{sanjay.jhingan@gmail.com} \emailAdd{djain@ddu.du.ac.in}

\abstract{The origin of the accelerated expansion of the Universe remains an unsolved mystery in
Cosmology. In this work we consider a spatially flat Friedmann-Robertson-Walker (FRW) Universe with
non-relativistic matter and a single scalar field contributing to the energy density of the Universe.
Properties of this scalar field, like potential, kinetic energy, equation of state etc. are
reconstructed from Supernovae and BAO data using Gaussian processes. We also reconstruct energy conditions and kinematic
variables of expansion, such as the jerk and the slow roll parameter. We find that the reconstructed scalar field variables and the kinematic quantities are consistent with a flat $\Lambda$CDM Universe. Further we find that the null energy condition is satisfied for the redshift range of the Supernovae data considered in the paper but, the strong energy condition is violated.}


\begin{document}

\maketitle

\section{Introduction}

Providing an explanation for the observed accelerated expansion of the Universe, is one of the bigger
challenges in Cosmology today. The evidence for existence of dark energy (DE) which is believed to
source this acceleration \cite{de,sah,sami}, has been accumulating ever since its first hints from
supernovae (SNeIa) observations \cite{sne}. Today there are multiple probes all substantiating the
idea that the Universe is accelerating \cite{accn}. However, even in this era of precision Cosmology,
we know little about this unknown form of energy. Many ambitious projects are being carried out and
others are lined up for the future with the goal of better understanding the nature of DE
\cite{surv}. On the theoretical side, the simplest and one of the most successful DE candidate is the
cosmological constant $\Lambda$, but it suffers from serious theoretical problems (fine tuning and
coincidence) \cite{wein}. If interpreted as a fluid with pressure and energy density, the
cosmological constant has an equation of state (eos) $\omega = -1$. Observations constrain the
eos of DE to be near $-1$ today, but the time variation in $\omega$ is still unconstrained. An
alternative to the cosmological constant are variable DE models. A variety of scalar field models
have been proposed in literature: quintessence, K-essence, Tachyon etc. \cite{sami}. All these models
are constructed within the framework of general relativity (GR). A modification of gravity at large
scales is also considered as a possible explanation for the late time acceleration of the Universe
(for a review see \cite{kunz} and references therein), but in this paper we work within the framework
of GR.

Various methods have been employed to understand the nature of DE by estimating DE parameters from
observations. The most frequently used model independent method is the parametric approach. A
standard way to model possible departure from the cosmological constant, is to parametrise the eos
parameter $\omega (z)$ to account for a value different from $-1$ (i.e. a possible variation with
redshift). One can either assume $\omega$ = constant or allow for specific redshift variations, for
example $\omega(z) = \omega _0 $+$ \omega_1 z/(1 + z)$, where $\omega_0$ and $\omega_1$ are
constants. Although this is a simple way to incorporate the dynamical aspects of DE and quantify its
properties with a few numbers, the results are strongly dependent on the parametrization chosen. For
example, Bassett et al., have shown that the confidence intervals inferred from standard
two-parameter expansions of the eos are typically untrustworthy \cite{bas}. Another way to
understand DE behaviour is to analyse kinematic variables like the Hubble rate $H(z)$, the
deceleration parameter $q(z)$ or the jerk parameter $j(z)$, which are all constructed from the
derivatives of the scale factor $a(z)$. There have been attempts to constrain $H_0$, $q_0$ and $j_0$
(the value of these parameters at $z$=0), by analysing the cosmographic expansion of the luminosity
distance \cite{vinc}. The kinematic approach is advantageous since one does not need to assume any
specific composition of the Universe and there have been attempts to reconstruct the cosmic expansion
history by parametrising $q(z)$ or $j(z)$ \cite{kine}. Also worth mentioning are the efforts to study
the geometric expansion history of the Universe, by Sahni and collaborators. They proposed the use of
diagnostic parameters called 'statefinders', constructed from the scale factor of the Universe and
its time derivatives \cite{sahni}. It is expected that future supernovae measurements would have the
constraining power to differentiate between different forms of dark energy using these diagnostic
parameters. In a related work, Sahni et al., proposed the {\it Om} diagnostic which is a combination of
the Hubble parameter and the cosmological redshift and provides a null test of dark energy being a
cosmological constant 
\cite{om}. This work was further extended by Shafieloo et al., who proposed the {\it Om3}
diagnostic \cite{om3sh}. This is designed for baryon acoustic oscillations (BAO) data and can combine information
from BAO and type Ia supernovae (SNeIa) to give another null diagnostic of the cosmological constant.

As mentioned earlier, parametrisation of DE variables can introduce bias in a analysis and
non-parametric techniques may offer a solution to these problems. Here one does not need to assume an
ad-hoc functional form for the quantities of interest ($\omega(z)$, $H(z)$, $q(z)$ or $j(z)$ etc.), which
could bias the results if the actual form of the evolution is different from the one assumed. In this
regard various non-parametric methods have been proposed, which use uncorrelated redshift binning,
Bayesian treatment based on maximum entropy methods etc. to reconstruct the expansion history and DE
eos \cite{ywang,trot}. Principal Component Analysis (PCA) is another (now) commonly used
non-parametric method. Huterer and Starkman were among the first to advocate the use of PCA to
analyse DE properties \cite{hut}. Shapiro and Turner used PCA to show that there is strong evidence
from the Riess supernovae data that the Universe once accelerated and that the deceleration parameter
was higher in the past \cite{shap} and Dick et al., presented a method for reducing the cosmological
data to constraints on the mode amplitudes of the dark energy density \cite{jason} using PCA. But it
is important to remember that this technique has caveats. Since most of the information is contained
in the first few eigenmodes, one makes a truncation and keeps the best constrained principal
components for further analysis. Keeping all the eigenmodes would amount to zero information loss,
but the error bars would be too large to give any meaningful estimates. The process of truncation can
potentially lead to biases and effect the estimated errors on the modes. Kitching and Amara showed
that the eigen-decomposition of the Fisher matrix (the main step of PCA) is sensitive to both the
order of expansion and the initial basis set used in the construction of the Fisher matrix
\cite{amara}. Putter and Linder further gave some quantified examples to argue that oversimplifying
PCA interpretation can lead to misinterpretations \cite{put}. 

Another non-parametric method, the
Smoothing technique developed by Shafieloo et al., uses Gaussian kernel to smooth the data (in their
case SNeIa data) to calculate cosmological functions like $H(z)$ and $\omega(z)$ \cite{arman1}. The technique can
also be used to distinguish between $\Lambda$CDM and evolving dark energy models.
Another technique, developed by Shafieloo, Clifton and Ferreira, the `Crossing statistics' can be
used even in cases where the intrinsic dispersion of a data set is not well known, and can
distinguish between different models in cases where the standard $\chi^2$ statistic fails
\cite{arman2}. In a recent paper Arman Shafieloo showed that these techniques can be combined to
reconstruct the expansion history of the Universe without putting any prior on the cosmological
quantities such as the eos \cite{arman}. Also worth mentioning is recent work by Nesseris and
Bellido who proposed a method for model independent and bias-free reconstruction of various
cosmological distances using Genetic algorithm \cite{nes}. They further showed that the method can be used to
estimate the errors on the reconstructed quantities analytically and their estimates agree with the
standard error estimation technique involving Fisher matrix.

An alternative non-parametric approach which has garnered attention lately is the Gaussian Process modelling (GP) which is the basis of this paper and is discussed in detail in the next section. GP is a non-parametric regression approach based on a generalization of the Gaussian probability distribution. Many attempts have been made to understand DE dynamics through GP methodology. GP has been used to reconstruct the eos of DE and to reconstruct the history of the expansion rate and deceleration parameter as a function of redshift~\cite{hols,seikel,shaf}. In a related work, similar in spirit to GP, Crittenden et al. developed a non-parametric Bayesian method for reconstructing the time evolution of the DE eos by choosing a prior which minimises variance and bias in the reconstruction \cite{fable}.

As mentioned earlier, one of the problems faced by the cosmological constant is the coincidence
problem (why DE dominates today?). It can be resolved by allowing the DE to be dynamical. Most of the
observational data is consistent with a $\Lambda$ cold dark matter ($\Lambda$CDM) Universe, and hence
any viable DE model has to mimic the cosmological constant today. Since there is no theoretical
compulsion for the DE density to be constant, and observations today cannot rule out dynamical
models for DE, there is a plethora of models which are proposed as an alternative to the cosmological
constant (for a review on dark energy reconstruction see \cite{sah}). In the presence of so many
theoretical models for DE, it becomes important to constrain the possible shape of the DE potential
to distinguish between different models. In this regard, there have been many attempts to reconstruct
DE properties. Saini et al., proposed a model-independent method, based on a versatile analytical
form for the luminosity distance for estimating the form of the potential of the scalar field driving
the cosmic acceleration, and the associated eos \cite{saini}. There have been several other
attempts to reconstruct the scalar potential and eos from observational data, see \cite{join} for examples.

The aim of this paper is to reconstruct and constrain the properties of the scalar field $\phi$,
which is assumed to contribute to the energy density of the Universe along with non-relativistic
matter. We also analyse energy conditions and two kinematic parameters, the `jerk' and the `slow
roll'. We consider a spatially flat FRW Universe since it agrees well
with the observations today. In section \ref{sec2}, we outline the main formulae used for the
reconstruction (\ref{mf}), discuss the GP methodology (\ref{gp}) and give details of the data sets
used (\ref{data}) for reconstruction. The results are presented and discussed in section
\ref{results}.

\section{Methodology}\label{sec2}
\subsection{Dark energy parameters}\label{mf}

{\bf Scalar field dynamics.}
We assume a flat FRW Universe with energy 
contribution coming from non-relativistic matter and a single scalar 
field. The equation governing the evolution of this scalar field is 
\begin{equation}
\ddot{\phi}+3H\dot{\phi}=-\frac{\partial V}{\partial \phi},
\end{equation}
and the pressure and density of the scalar field (assuming it is 
spatially homogeneous) is given by 
\begin{align}
P_\phi &= \frac{\dot{\phi}^2}{2} -V(\phi), \\
\rho _\phi &= \frac{\dot{\phi}^2}{2} +V(\phi).
\end{align}
The Friedmann equations in this case reduce to 
\begin{align}
H^2 &= \frac{8\pi G}{3}(\rho _m + \rho _\phi), \\
\frac{\ddot{a}}{a} &= -\frac{4 \pi G}{3} (\rho _m + \rho _\phi + 3 P_\phi).
\end{align}
Re-arranging the above two sets of equations we get,
\begin{align}
\frac{8 \pi G}{3H^2_0} ~ V(z) &= \frac{H^2}{H^2_0}-\frac{1+z}{3H^2_0}~H H'-
								\frac{\Omega _m (1+z)^3}{2},
\label{eqnv} \\  
\frac{8 \pi G}{3H^2_0} ~ \dot{\phi}^2 &= \frac{2(1+z)}{3H^2_0}~H H'
										-\Omega _m (1+z)^3.
\label{eqnphi}
\end{align}
Here, an over-dot indicates differentiation with respect to time, and 
prime indicates differentiation with respect to redshift $z$. 
Similarly, one can reconstruct the eos of the scalar field $\omega$ (=$P_{\phi}$/$\rho_{\phi}$), using these expressions. Recent constraints on a constant eos from Planck collaboration 
is $\omega=-1.09\pm 0.17$ (Planck+WMAP+Union2.1) and on a two 
parameter model of eos $\omega(z)=\omega_0+\omega_a(1-a)$ is 
$\omega_0 = -1.04^{+0.72}_{-0.69}$ and $\omega_a<1.32$ 
(all error bars are at 95\% confidence) \cite{planck}.

The evolution of these scalar field variables ($V$, $\dot{\phi}^2$, $\omega$)
can be estimated, if one can reconstruct the Hubble parameter and its derivatives from observations. Now, in a flat FRW Universe the luminosity  distance $D_L(z)$, can be written as 
\begin{equation}
D_L(z)= \frac{c(1+z)}{H_0} \int_{0}^{z} \frac{dz'}{{\cal H}(z')}, 
\end{equation}
where $c$ is the speed of light, $H_0=H(z=0)$ and ${\cal H}(z)$=$H(z)$/$H_0$. Since the Hubble parameter can be written in terms of the derivative of the distance, we can also derive scalar field variables in terms of $D_L(z)$ (and its derivatives). \\

\noindent{\bf Energy conditions.} In general relativity (GR), ``energy conditions" is a way of implementing the experimentally observed notion of positivity of energy density. Different energy conditions impose restrictions on various linear combinations of energy momentum tensor. The powerful theorems proven in GR, the singularity theorems, the positive energy theorem etc., all assume some notion of positivity of energy density implemented through energy conditions. These conditions were introduced in Hawking and Ellis \cite{HE} as coordinate independent inequalities on energy momentum tensor in Einstein field equations. Because of their simplicity and model independence they are widely discussed in GR \cite{wald}. Various energy conditions (null energy condition (NEC), weak energy condition (WEC), strong energy condition (SEC), and dominant energy condition (DEC)), in the setting of FRW background can be expressed as constraints on energy density and pressure as \cite{caroll}
\begin{eqnarray}
NEC: &\rho + p \geq 0, &  \nonumber \\
WEC: &\rho +p \geq 0, \quad \text{and} & \rho \geq 0 \nonumber \\
SEC: &\rho + p \geq 0,  \quad \text{and} & \rho + 3 p \geq 0 \nonumber \\
DEC: &\rho \geq 0,  \quad \text{and} & -\rho \leq p \leq p 
\end{eqnarray}
The energy conditions and their cosmological relevance was first explored in detail by Matt Visser \cite{matt}. Later on, various authors used observational data to study the violations of the energy conditions and their consequences \cite{ec}. Thus energy conditions play an important role in imposing model independent bounds on some of the key properties of DE.
In this paper we explore the validity of the null and strong energy conditions (NEC \& SEC). The SEC
implies that the expansion of the Universe is slowing down, hence violation of SEC ($\ddot{a}>0$)
means accelerated expansion. The violation of SEC also implies $\omega \equiv P/\rho < -1/3$. \\

\noindent{\bf Kinematic parameters.} 
In the absence of a compelling model to describe the cosmic 
acceleration, it is useful to study kinematic parameters derived 
from the derivatives of the scale factor $a$. The Hubble parameter 
$H(t) = \dot{a}/a$, the deceleration parameter 
$q(t) = -a\ddot{a}/\dot{a}^2$, and the jerk parameter 
$j(t) = \dddot{a}a^2/a^3$, are such kinematic parameters and are all theoretically equivalent since one can be written in 
terms of the derivative or integral of the other. The jerk parameter is favored sometimes as a good kinematic choice, since for a $\Lambda$CDM Universe $j(z)=1$, for all $z$. Hence it can be used to distinguish different DE models and there have been several attempts to constrain the jerk parameter using observational data \cite{rap,lima,zhai}. $j$ can be expressed in terms of the Hubble parameter as:
\begin{equation}
j = 1+\frac{(1+z)^2
  H''(z)}{H(z)}-\frac{2(1+z)H'(z)}{H(z)}+\frac{(1+z)^2
  H'(z)^2}{H(z)^2}.
\end{equation}

The Slow roll parameter $\epsilon_H=-\ddot{a}/aH^2+1$ is another parameter employed to understand the
dynamics of expansion due to the scalar field and can be used to check the presence of an
inflationary phase ($\epsilon_H<<1$) in the Universe. Note that in the case of late time DE dynamics
this parameter is not entirely reliable as a scalar field parameter, since the density of matter is
not negligible. But one can define this parameter in terms of the derivative of the Hubble parameter
and use it as a measure to check the existence of an accelerated expansion \cite{sami}. $\epsilon_H <
1$ is equivalent to the accelerated expansion $\ddot{a} > 0$, and $\epsilon_H > 1$ is the decelerated
expansion $\ddot{a} < 0$, where 
\begin{equation} 
	\epsilon_H = (1+z)\frac{H'(z)}{H(z)}. 
\end{equation}
Under the slow roll limit ($\epsilon_H << 1$) the Hubble parameter becomes almost constant and the
Universe expands as if it is dominated by the cosmological constant.

\subsection{Gaussian Processes}\label{gp}

A Gaussian Process (GP) is a collection of random variables, any finite number of which have a joint Gaussian distribution \cite{rasm}. Similar to a Gaussian distribution, which is a distribution of a random variable characterized by a mean and a covariance, a GP is a distribution over functions, characterized by a mean function and a covariance matrix. 
Given a set of observations the aim is to infer the relation between independent and dependent variables. In parametric regression one assumes some functional dependence of the output on the input $f(x,{\boldsymbol \theta})$, where ${\boldsymbol \theta}$ represents the set of model parameters. Here regression entails finding the values of ${\boldsymbol \theta}$ which best describe the data. The best fit parameters are usually found by minimizing a chi-squared merit function. Similarly, in GP modelling the function $f(x)$ is represented as 
\begin{equation} 
f(x) \sim GP(\mu (x),k(x,x')),
\end{equation}
i.e, the value of $f(x)$ at any point $x$, is a Gaussian random variable with mean $\mu (x)$ and covariance $k(x,x')$:
\begin{eqnarray}
\mu (x) &=& E(f(x)),\\
k(x,x') &=& E((f(x)-\mu (x))(f(x')-\mu (x'))).
\end{eqnarray}
A number of possible choices for the covariance functions exist: squared exponential, polynomial, spline etc. By choosing a particular covariance function, we choose a basis of the function space to expand the target function. Here we chose the commonly used squared exponential function for its simplicity and also because it is infinitely differentiable, a property which is useful while constructing higher order 
derivatives. The squared exponential covariance function is expressed as:
\begin{equation}
k(x,x')=\sigma_f^2 \exp \left(-\frac{(x-x')^2}{2l^2}\right).
\end{equation}
This functions is parameterized by the two parameters, 
$\sigma_f$ 
and $l$ (known as hyperparameters). These parameters represent the length scales in the GP.
$l$ corresponds to the correlation length along which the successive $f(x)$ values are 
correlated and $\sigma_f$ controls the variation in $f(x)$ relative to 
the mean of the process. Thus, the covariance between output variables are written as a function of the inputs. One can see that the covariance is maximum for variables whose inputs are very close. This is something we do expect for smooth functions. The matrix elements of the covariance matrix for the GP: $K({\bf X,X})$,
are given by 
\begin{equation}
[{\bf K(X,X})]_{i,j}=k(x_i,x_j)
\end{equation}
One can generate random functions using this covariance matrix even without any data. 
Similar to the function $f(x)$, the data ${\bf y}$ can be represented using GP:
\begin{equation} 
y \sim GP(\mu (x),k(x,x')),
\end{equation} 
Given a set of inputs ${\bf X}$ (also called training vectors), outputs ${\bf y}$ (the data set, also called target) 
and the covariance matrix ${\bf K(X,X})$, our aim is to make inference about the function $f(x)$ at some other points ${\bf \hat{X}}$. The joint probability distribution for data $y$ and reconstructed function $\hat{f}$ is given by 
\[ 
\begin{pmatrix} 
{\bf y} \\ {\bf \hat{f}} 
\end{pmatrix}  \sim  N \left( \begin{bmatrix} {\boldsymbol \mu} \\ \hat{{\boldsymbol \mu}} \end{bmatrix}, 
\begin{bmatrix} {\bf K(X,X)+C} & {\bf K(X,\hat{X}})\\ {\bf K(\hat{X},X}) &
  {\bf K(\hat{X},\hat{X}}) \end{bmatrix} \right), \]  
where ${\boldsymbol \mu}$ and $\hat{{\boldsymbol \mu}}$ are the assumed means (initial guess) and ${\bf C}$ is the covariance matrix of the data. If the data points are uncorrelated then ${\bf C}$ is a diagonal
matrix. Say, $p$ and $q$ are the number of points in ${\bf X}$ and ${\bf \hat{X}}$ respectively, and the joint covariance matrix
\[
{\bf \Sigma} = 
\begin{bmatrix}
{\bf K(X,X)+C}, & {\bf K(X,\hat{X}})\\ 
{\bf K(\hat{X},X}), & {\bf K(\hat{X},\hat{X}})
\end{bmatrix}
=
\begin{bmatrix}
{\bf \Sigma}_{11} & {\bf \Sigma}_{12}\\ {\bf \Sigma}_{21} & {\bf \Sigma}_{22}
\end{bmatrix},\]
then one can write
\[
P({\bf y},{\bf \hat{f}}) = \frac{1}{(2\pi)^{(p+q)/2} |{\bf \Sigma}|^{1/2}} \exp \left\lbrace \begin{bmatrix}
({\bf y}-{\boldsymbol \mu})^T, & ({\bf \hat{f}}-\hat{{\boldsymbol \mu}})^T 
\end{bmatrix}
{\bf \Sigma}^{-1}
\begin{bmatrix}
{\bf y}-{\boldsymbol \mu}\\ {\bf \hat{f}}-\hat{{\boldsymbol \mu}}
\end{bmatrix} \right\rbrace
\]
The matrix elements of the inverse of the joint covariance 
matrix ${\bf \Sigma}$ can be evaluated as follows
\[
{\bf \Sigma}^{-1} = 
\begin{bmatrix}
{\bf K(X,X)+C}, & {\bf K(X,\hat{X}})\\ 
{\bf K(\hat{X},X}), & {\bf K(\hat{X},\hat{X}})
\end{bmatrix}^{-1}
=
\begin{bmatrix}
{\bf \Sigma}^{11} & {\bf \Sigma}^{12}\\ {\bf \Sigma}^{21} & {\bf \Sigma}^{22}
\end{bmatrix}\]
where 
\begin{multline*}
{\bf \Sigma}^{11}={\bf (K(X,X)+C)}^{-1}+{\bf (K(X,X)+C)}^{-1} {\bf K(X,\hat{X})} \\
[{\bf K(\hat{X}},{\bf \hat{X}})-{\bf K(X,\hat{X})}^T ({\bf K(X,X)+C})^{-1} {\bf K(X,\hat{X}})]^{-1} \\
{\bf K(X,\hat{X}})^T ({\bf K(X,X)+C})^{-1}
\end{multline*}
\begin{multline*}
{\bf \Sigma}^{22}={\bf K(\hat{X},\hat{X}})^{-1}+{\bf K(\hat{X},\hat{X}})^{-1} {\bf K(X,\hat{X}})^T \\
[{\bf K(X,X)+C}-{\bf K(X,\hat{X}}) {\bf K(\hat{X},\hat{X}})^{-1} {\bf K(X,\hat{X}})^T]^{-1} \\
{\bf K(X,\hat{X}}) {\bf K(\hat{X},\hat{X}})^{-1}
\end{multline*}
\begin{multline*}
{\bf \Sigma}^{12}=-({\bf K(X,X)+C})^{-1} {\bf K(X_,\hat{X}}) [{\bf K(\hat{X},\hat{X}})-{\bf K(X,\hat{X}})^T- \\
({\bf K(X,X)+C})^{-1} {\bf K(X,\hat{X}})]^{-1}
\end{multline*}
Further the determinant of the matrix ${\bf \Sigma}$ is given by 
\[
det({\bf \Sigma})=det({\bf \Sigma}_{11})~~det({\bf \Sigma}_{22}-{\bf \Sigma}_{12}^T {\bf \Sigma}_{11}^{-1} {\bf \Sigma}_{12})
\]
Substituting these expressions for ${\bf \Sigma}^{11}$, ${\bf \Sigma}^{12}$ \& ${\bf \Sigma}^{22}$ 
we obtain the joint probability distribution for 
data ${\bf y}$ and the reconstructed function ${\bf \hat{f}}$ as
\begin{multline*}
P({\bf y},{\bf \hat{f}})=\frac{1}{(2\pi)^{p/2} |{\bf \Sigma}_{11}|^{1/2}} \exp [-\frac{1}{2}({\bf y}-{\boldsymbol \mu})^T {\bf \Sigma}_{11}^{-1}({\bf y}-{\boldsymbol \mu})]\\
\frac{1}{(2\pi)^{q/2} |{\bf A}|^{1/2}} \exp [-\frac{1}{2}({\bf \hat{f}}-{\bf a})^T {\bf A}^{-1}({\bf \hat{f}}-{\bf a})],
\end{multline*}
where 
\[
{\bf a}={\hat{\boldsymbol \mu}}+{\bf K(X,\hat{X}})^T ({\bf K(X,X)+C})^{-1}({\bf y}-{\boldsymbol \mu})
\]
and
\[
{\bf A}={\bf K(\hat{X},\hat{X}})-{\bf K(X,\hat{X}})^T ({\bf K(X,X)+C})^{-1} {\bf K(X,\hat{X}})
\]
The marginal distribution of ${\bf y}$ 
is given by
\begin{equation}
P({\bf y})=\int P({\bf y},{\bf \hat{f}}) d{\bf \hat{f}} =\frac{1}{(2\pi)^{p/2} |{\bf \Sigma}_{11}|^{1/2}} \exp [-\frac{1}{2}({\bf y}-{\boldsymbol \mu})^T {\bf \Sigma}_{11}^{-1}({\bf y}-{\boldsymbol \mu})],
\label{margp}  
\end{equation}
and the conditional distribution $P({\bf \hat{f}}|{\bf y})$ is
\[
P({\bf \hat{f}}|{\bf y})=\frac{P({\bf y},{\bf \hat{f}})}{P({\bf y})}=\frac{1}{(2\pi)^{q/2} |{\bf A}|^{1/2}} \exp [-\frac{1}{2}({\bf \hat{f}}-{\bf a})^T {\bf A}^{-1}({\bf \hat{f}}-{\bf a})],
\]
i.e, the reconstructed function ${\bf \hat{f}}({\bf {\hat X}})$ has a Gaussian normal distribution given by
\begin{equation}
{\bf \hat{f}} \sim GP({\bf a},{\bf A}).
\end{equation}
Note that $\sigma_f$ and $l$ are unknown parameters of the GP. The training of a GP involves selecting these parameters. 
This can be done by maximising the marginal log-likelihood probability $\ln P(y)$ (from \ref{margp})
\[
\ln P({\bf y})= - \frac{1}{2} ({\bf y}-{\boldsymbol \mu})^T [{\bf K(X,X)+C}]^{-1} ({\bf y}-{\boldsymbol \mu}) -\frac{1}{2}
\ln |{\bf K(X,X)+C}|-\frac{p}{2} \ln 2 \pi 
\]
This is an approximation, and can be used if the posterior for ${\boldsymbol \theta }$ (which includes any other parameters that the 
analysis may have), is fairly well peaked. If there are many parameters in the analysis (which include those coming from an assumed mean function), this kind of optimization may result in over-fitting. In our analysis, we sample the hyperparameter space and the probability distributions of the reconstructed function are weighted by the posterior distributions of the hyperparameters. We also need to reconstruct derivative function for our analysis.
The derivative of a GP is another GP and hence, GP modelling can be used 
to reconstruct the derivatives as well. If $k(x_i,x_j)$ is the covariance 
between function values then the covariance between function and its 
derivatives, and between derivatives can be obtained by differentiating 
the original covariance function
\[
cov\left(f_i, \frac{\partial f_j}{\partial x_j}\right)= \frac{\partial
  k(x_i,x_j)}{\partial x_j}, ~~ 
cov\left(\frac{\partial f_i}{\partial x_i}, \frac{\partial
  f_j}{\partial x_j}\right)= \frac{\partial^2 k(x_i,x_j)}{\partial
  x_i \partial
  x_j}, 
\] 
and one can evaluate the covariance matrices for the derivatives:\\ $[{\bf K'(X,\hat{X}})]_{i,j}$ = $\partial k(x_i,\hat{x}_j)$ / $\partial
\hat{x}_j$, $[{\bf K''(\hat{X},\hat{X}})]_{i,j}$ = $\partial^2 k(\hat{x}_i,\hat{x}_j)$ / $\partial \hat{x}_i \partial \hat{x}_j$ etc.\\As before, the mean value and the covariance for the reconstructed derivative function can be obtained from the conditional distribution $P({\bf \hat{f'}}|{\bf y})$ as follows (and higher order derivatives can be constructed similarly),
\begin{equation}
\bar{{\bf \hat{f'} }}=\hat{{\boldsymbol \mu'}}+{\bf K'(X,\hat{X}})^T ({\bf K(X,X)+C})^{-1}({\bf y}-{\boldsymbol \mu})
\end{equation}
and
\begin{equation}
cov({\bf \hat{f'}})={\bf K''(\hat{X},\hat{X}})-{\bf K'(X,\hat{X}})^T ({\bf K(X,X)+C})^{-1} {\bf K'(X,\hat{X}}).
\end{equation}
To construct functions that depend on combinations of these reconstructed functions i.e. ${\bf \hat{f}}$, ${\bf \hat{f'}}$ ${\bf \hat{f''}}$ etc., one samples from the joint probability distributions of these functions.
\begin{figure}[ht]
\centering \subfloat[Part 3][]
{\includegraphics[width=3in]{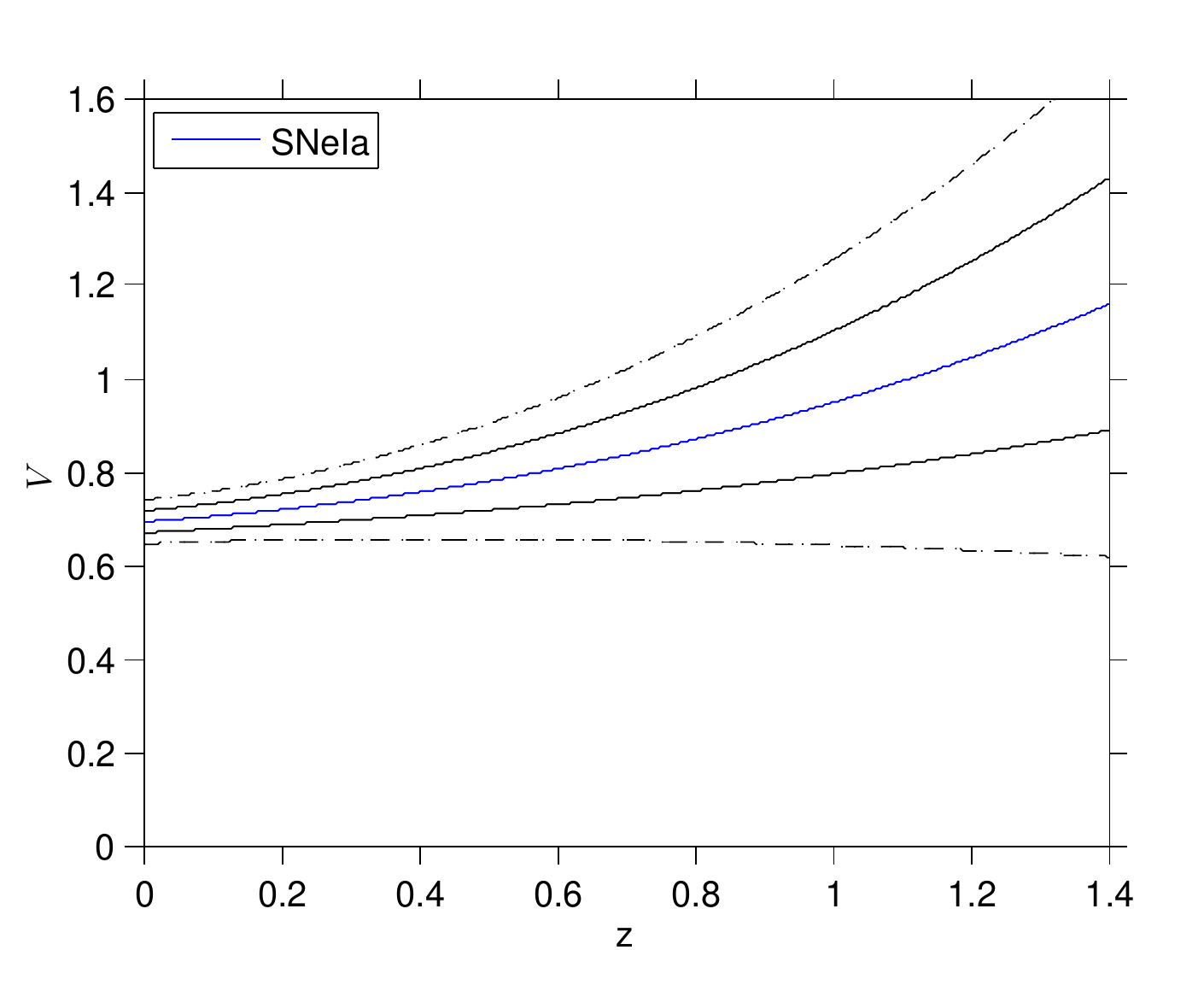}\label{fig12a}} 
\subfloat[Part3][]
{\includegraphics[width=3in]{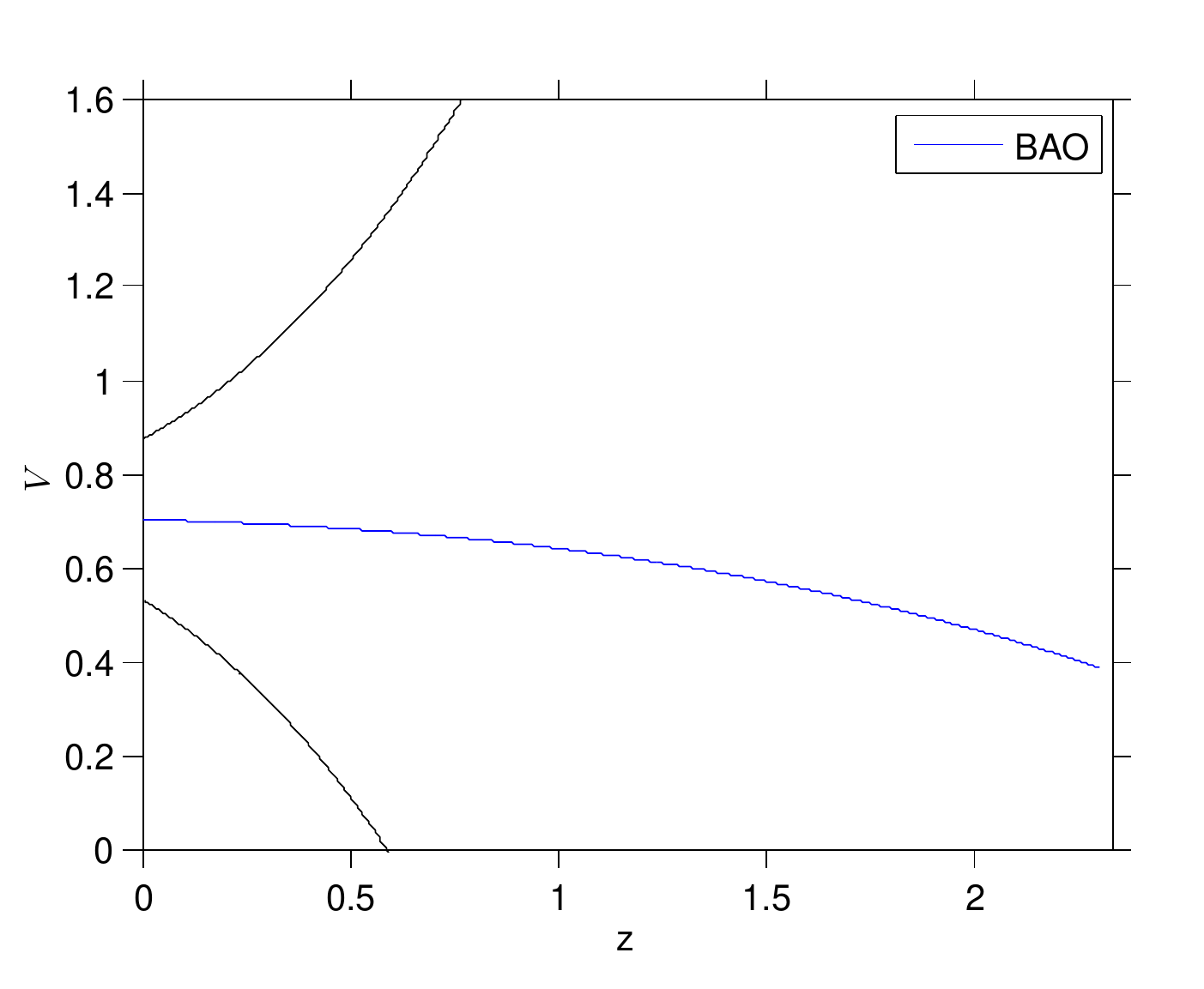}\label{fig13a}} 
\caption{Solid lines show the reconstructed potential as a function of redshift with 1$\sigma$ error bars (black curves). In figure (a) the dashed lines corresponds 2$\sigma$ error bars.}
\label{figaa}
\end{figure}
\begin{figure}[ht]
\centering \subfloat[Part 3][]{
\includegraphics[width=3in]{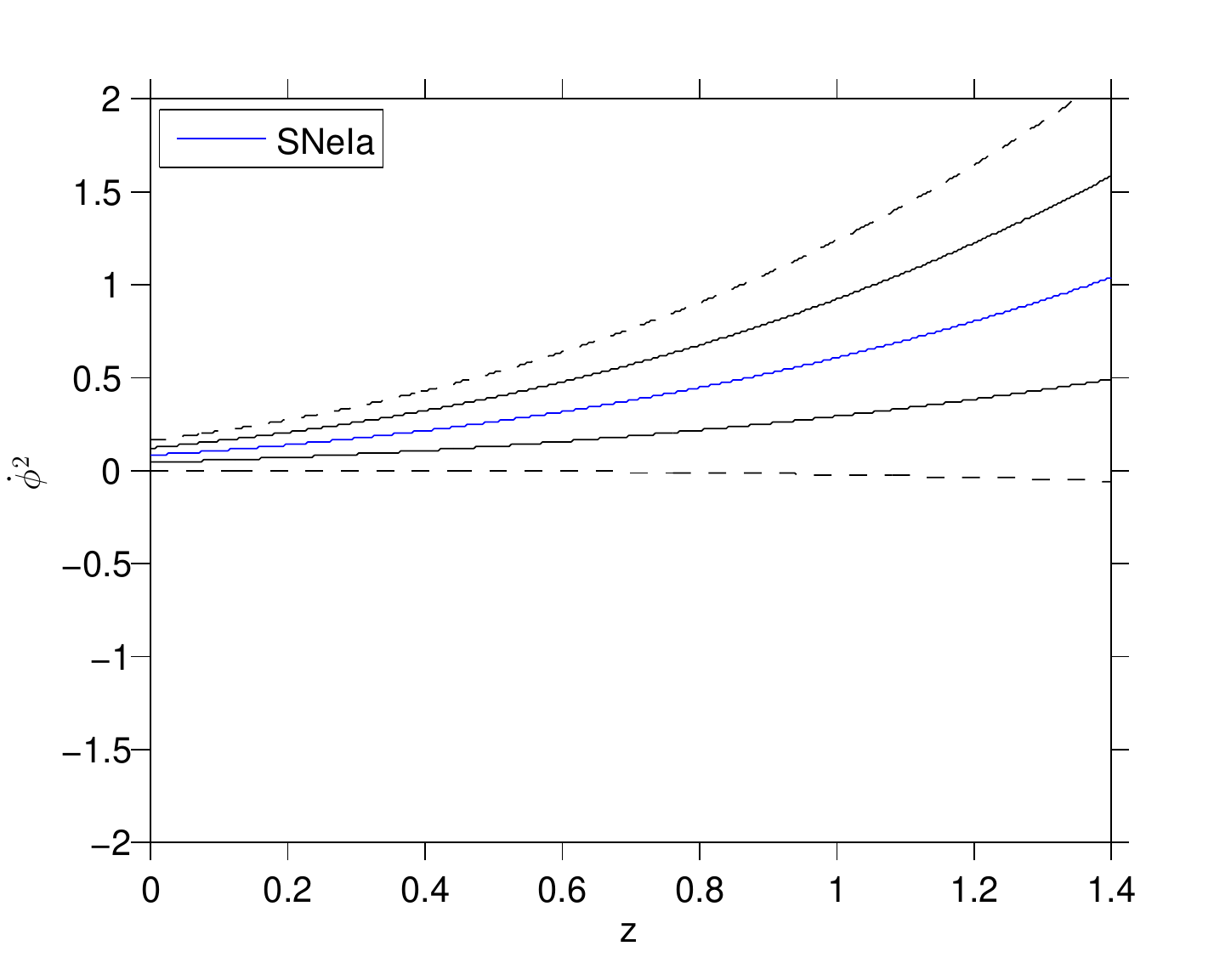}\label{fig21a}} \subfloat[Part
3][]
{\includegraphics[width=3in]{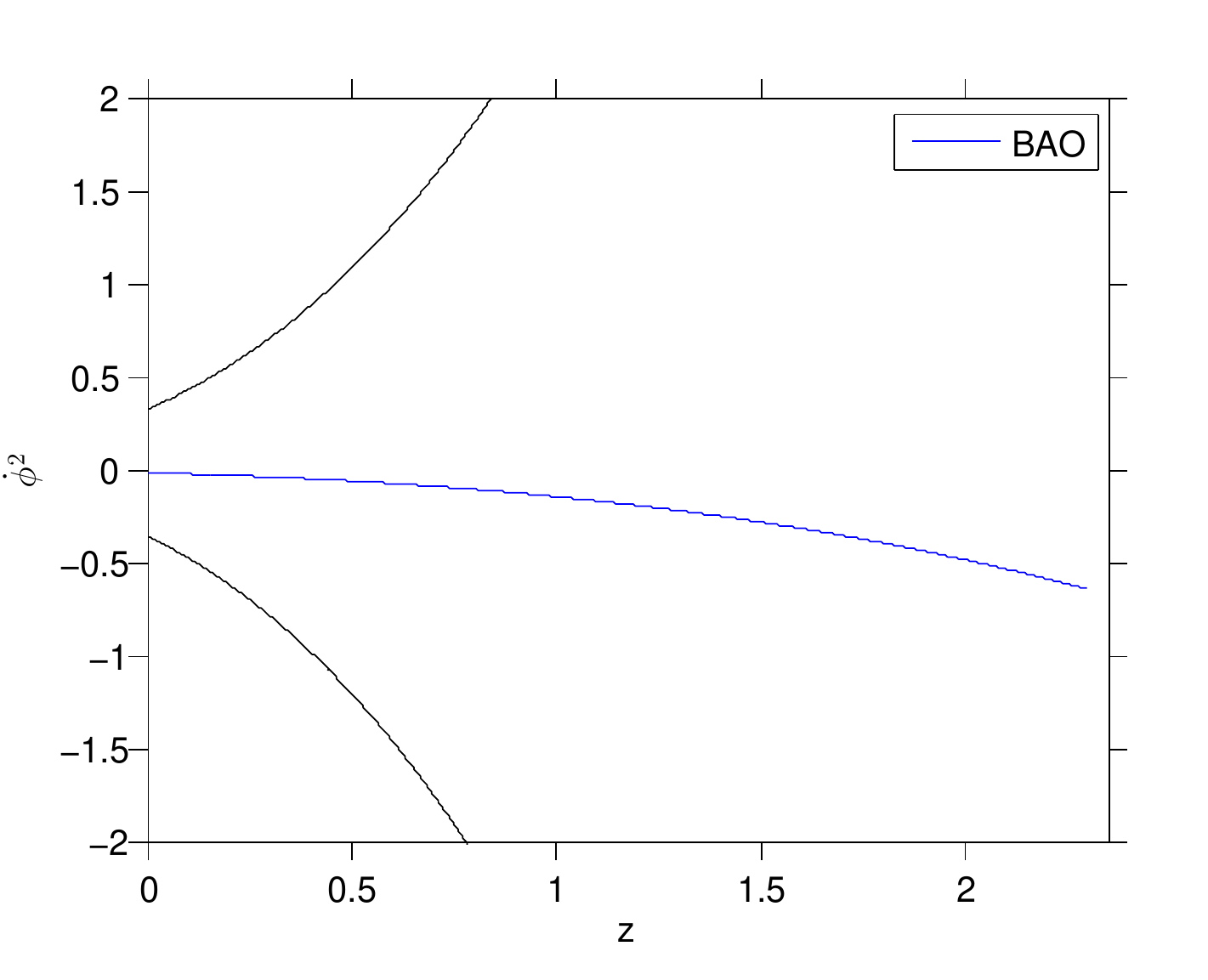}\label{fig22a}} 
\caption{Solid lines show the reconstructed $\dot{\phi}^2$ as a function of redshift with 1$\sigma$ error bars (black curves). In figure (a) the dashed lines corresponds 2$\sigma$ error bars.}  
\label{figba}
\end{figure}
\subsection{Data sets}\label{data}
For our analysis we used distance modulus data from SNeIa and 
Hubble measurements from BAO data. The priors on all the hyperparameters are assumed to be flat, and we give rather broad prior ranges as specified below for each data set. The details of the data sets used are given below:\\

\noindent {\bf BAO data.} 
We use eight $H(z)$ measurements from BAO data 
compiled from different groups in the redshift range $0.24<z<2.3$.
Gaztanaga et al., used the spectroscopic Sloan Digital Sky Survey (SDSS)
data to give a direct measurement of 
the Hubble parameter $H(z)$ as a function of redshift at $z=0.24$ 
\& $0.43$ \cite{gazt}. Chuang and Wang used the sample of luminous 
red galaxies from the SDSS Data Release 7 to give Hubble measurement 
at $z=0.35$ \cite{wang}. Blake et al., gave measurements of $H(z)$ at 
redshifts $z = 0.44, 0.6$ and $0.73$ by combining measurements of 
the baryon acoustic peak and Alcock-Paczynski distortion from galaxy 
clustering in the WiggleZ Dark Energy Survey \cite{blake}. 
Kazin et al., analysed the SDSS-III Baryon Oscillation Spectroscopic 
Survey (BOSS) CMASS sample of massive galaxies of the ninth data release 
to measure $H(z)$ at redshift $z=0.57$ \cite{kazin}. Finally, 
Busca et al., used the three-dimensional correlation function of the 
Lyman $\alpha$ forest of high-redshift quasars from the SDSS-III BOSS 
sample to provide a measurements of $H(z)$ at $z=2.3$ \cite{busca}. The data is used to reconstruct $H(z)$ and higher order derivatives simultaneously using GP, and then these functions are used to estimate the dark energy variable. For reconstruction, the initial guess for the mean function corresponds to flat $\Lambda$CDM Universe. There are four hyperprameters in this case: $\sigma_f$ and $l$, (from the choice of the covariance function) and $H_0$ and $\Omega_m$ (from the choice of the mean function). The priors on these parameters are: 0.002 $<\sigma_f<$ 1, 0.0001 $<l<$ 2, 50 $<H_0<$ 80 and 0.1 $<\Omega_m<$ 0.6. The likelihood for this data set can be written as:
\begin{equation*}
L({\boldsymbol \theta}) \propto   \frac{1}{\prod_i \sigma_i} \exp\left(- \frac{1}{2} \sum_i \left({\frac{H_i - H(z_i,{\boldsymbol \theta})}{\sigma_i}}\right)^2 \right), 
\end{equation*}
where $\boldsymbol \theta$ encapsulated the hyperparameters mentioned above and $\sigma_i$ is the error on each $H(z_i)$ as given in the data.
\begin{figure}[ht]
\centering \subfloat[Part 3][]{
\includegraphics[width=3in]{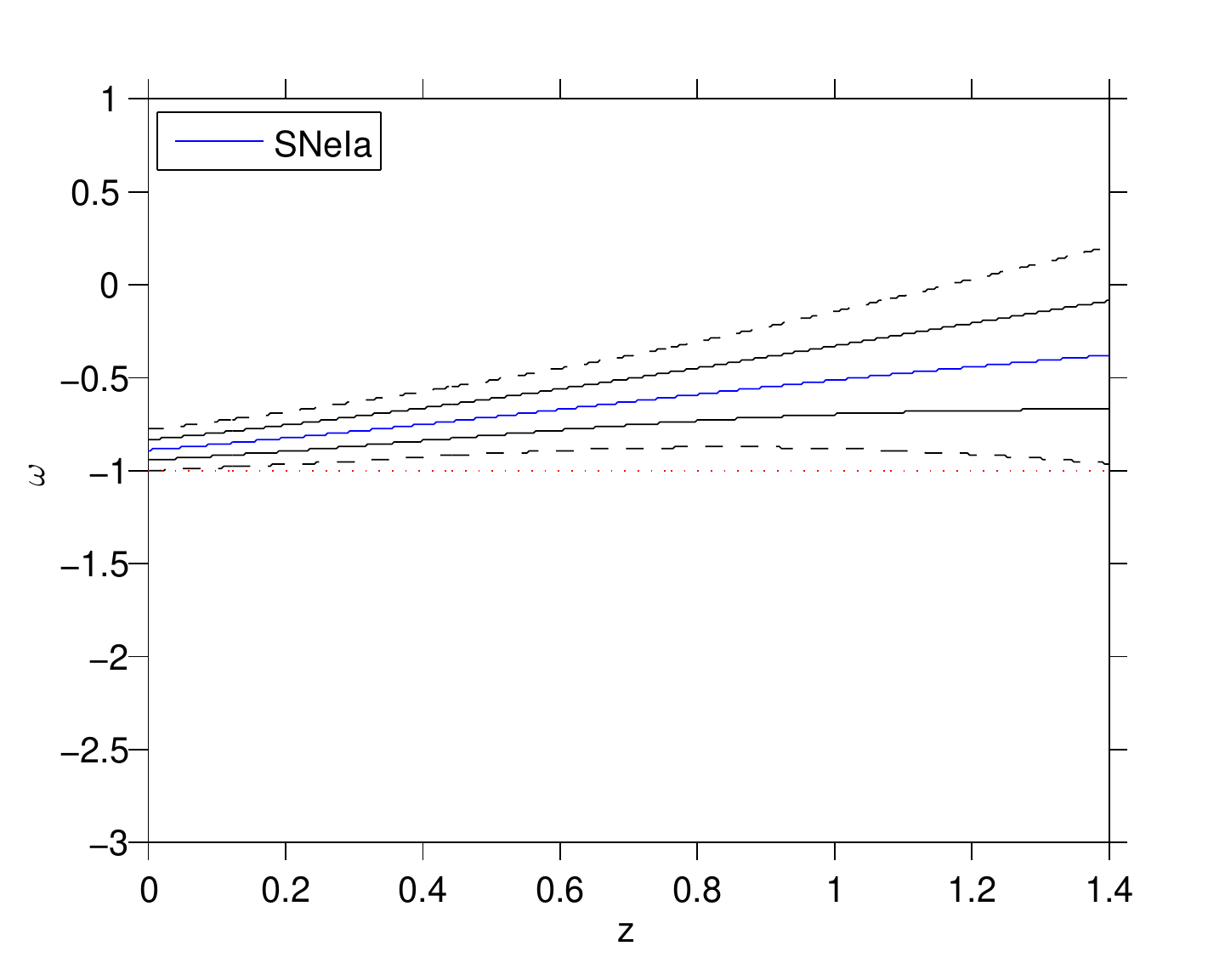}\label{fig51a}} \subfloat[Part
3][]
{\includegraphics[width=3in]{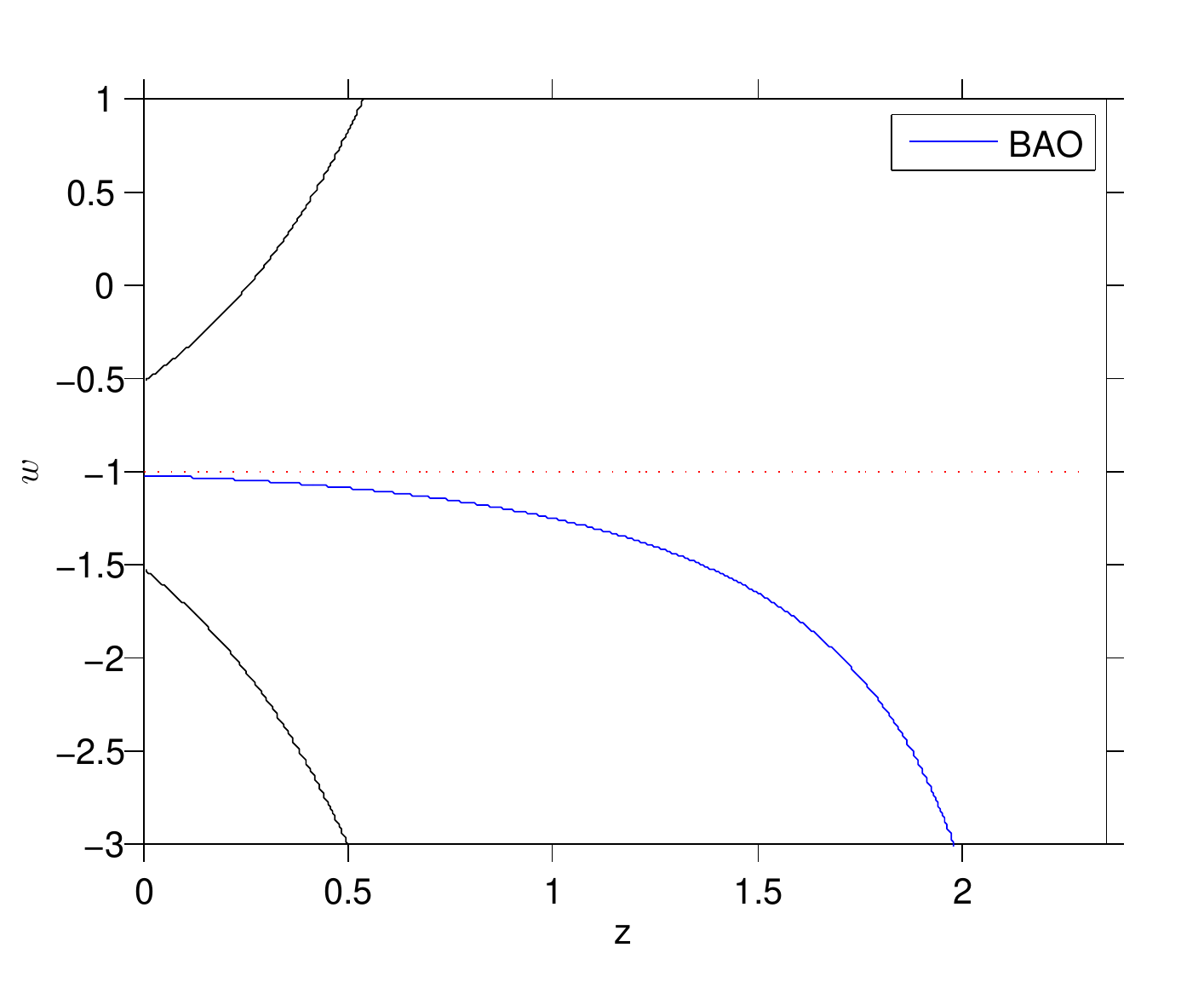}\label{fig52a}} 
\caption{Solid lines show the reconstructed eos as a function of redshift with 1$\sigma$ error bars (black curves). In figure (a) the dashed lines corresponds 2$\sigma$ error bars.}  
\label{figea}
\end{figure}

\noindent{\bf Supernovae data.}
We use distance modulus measurements from SNeIa Union2.1 sample, as described in 
\cite{suz}, to estimate the luminosity distance $D_L(z)$. This sample 
contains 580 supernovae spanning the redshift range $0.015 < z < 1.414$. The distance modulus in terms of the redshift, as given in the data is used to estimate the luminosity distance. The relation between the distance modulus $\mu$ and the luminosity distance $D_L$ is:
\begin{equation}
\mu_B(z) = m_B - M_B = 5 \log_{10} \left(\frac{D_L(z)}{1Mpc}\right)+25, 
\end{equation}
where $M_B$ is the absolute magnitude of the source and $m_B$ is the apparent magnitude ($B$ is for $B$-band). One major source of systematic uncertainty here is the unknown absolute magnitude $M_B$. Following Holsclaw et al., we define the distance measure $D(z)=H_0D_L(z)$ and use it in our analysis (see \cite{hols} for details). Here the re-defined distance modulus $\mu'$ is 
\begin{equation}
\mu'(z) = \mu_B(z)+5\log_{10}(H_0)-25=5\log_{10}(D(z)).
\end{equation}
The uncertainties in the calibration are accounted for by introducing a shift parameter $\Delta_{\mu}$, in the analysis. The Hubble rate can be written in terms of $D(z)$ as  
\begin{equation*}
H(z) = \frac{cH_0(1+z)^2}{D'(z)(1+z)-D(z)}. 
\end{equation*}
We can write equations (\ref{eqnv}) and (\ref{eqnphi}) in terms of $D(z)$, and its derivatives as follows:
\begin{align}
\frac{8\pi G}{3H_0^2}V(z) &= \frac{c^2(1+z)^4}{3(D'(z)(1+z)-D(z))^2} \left[1+\frac{D''(z)(1+z)^2}{D'(z)(1+z)-D(z)} \right]-\frac{\Omega_m }{2}(1+z)^3 ,
\\
\frac{8\pi G}{3H_0^2}\dot{\phi}^2 &= \frac{2c^2(1+z)^4}{(D'(z)(1+z)-D(z))^2} \left[2-\frac{D''(z)(1+z)^2}{D'(z)(1+z)-D(z)}\right]-\Omega_m(1+z)^3.
\end{align} 
Other dark energy variables like the pressure, energy density, equation of state etc. can be derived using the above formulae. Similarly the kinematic variables like $\epsilon_H$ and the jerk $j$ are derived in terms of $D(z)$ (and its derivatives),
\begin{eqnarray*}
\epsilon_H &=& 2 - (1+z)^2 \frac{D''(z)}{D'(z)(1+z)-D(z)},\\
j &=& 3 - \frac{(1+z)^2}{{(D'(z)(1+z)-D(z))^2}} \left[7 D''(z) + (1+z) D'''(z) - 3 \frac{(1+z)^2 D''(z)^2}{(D'(z)(1+z)-D(z))}\right].
\end{eqnarray*}
The GP method is used to reconstruct $D(z)$ and higher order derivatives simultaneously, and then these are used to derive the dark energy variables. For reconstruction, the initial guess for the mean function corresponds to flat $\Lambda$CDM Universe. There are 5 hyperprameters in this case: $\sigma_f$ and $l$, (from the choice of the covariance function) and $\Omega_m$, $\Delta_{\mu}$ and $\tau^2$ (from the choice of the mean function and overall scaling of the data). Here $\tau$ accounts for a possible rescaling \cite{hols}. The priors on these parameters are: 0.002 $<\sigma_f<$ 1, 0.0001 $<l<$ 1.4, 0.1 $<\Omega_m<$ 0.6, -0.5 $<\Delta_{\mu}<$ 0.5 and 0.5 $<\tau^2<$ 1. The likelihood in this case can be written as:
\begin{equation*}
L({\boldsymbol \theta}) \propto   \frac{1}{\tau^n \prod_i\sigma_i} \left[\exp\left(- \frac{1}{2} \sum_i \left({\frac{D_i - D(z_i,{\boldsymbol \theta})}{\sigma_i \tau}} \right)^2 \right)\right], 
\end{equation*}
where $\boldsymbol \theta$ encapsulates the hyperprameters mentioned above.

\section{Results and discussion}\label{results}
We scan the hyperparameter space using MCMC sampling for both data sets. The probability distribution of the reconstructed function $H(z)$, $H'(z)$ and $H''(z)$, and $D(z)$, $D'(z)$, $D''(z)$ and $D'''(z)$, are weighted by the posterior distribution of the hyperparameters (since our priors are flat, the posterior of the hyperparameters are proportional to the likelihood). For the reconstructed function we obtain a sum of the weighted Gaussian distributions as a result of the marginalization (the integral in the marginalisation of the hyperparameters is approximated as a sum in the discrete case). The mean and covariances for these mixture distributions can be evaluated analytically since they have closed forms (i.e. they too are Gaussian). These weighted mean functions and covariances are then used to derive the dark energy variables and the kinematic variables of cosmic expansion. Note here that variables like the potential, kinetic energy, equation of state etc. have explicit dependence on  $\Omega_m$. In such cases we use the mean value of $\Omega_m$, as obtained from the posterior hyperparameter distribution and include its variance in the overall error. 

We also note here that our results may be effected by the choice of the initial mean function ($\Lambda$CDM in our case). In this regard, a better choice could be to incorporate an iteration procedure in the analysis, where the posterior found from some preliminary runs are used as mean functions of successive runs (similar to \cite{hols}). Another method is to adopt some smoothing procedures to generate an initial guess of the mean from say multiple functions \cite{shaf}, and then iterate this to obtain the final results. These methods may give more robust results which are independent of the initial mean function. 
\begin{figure}[ht]
\centering \subfloat[Part 3][]{
\includegraphics[width=3in]{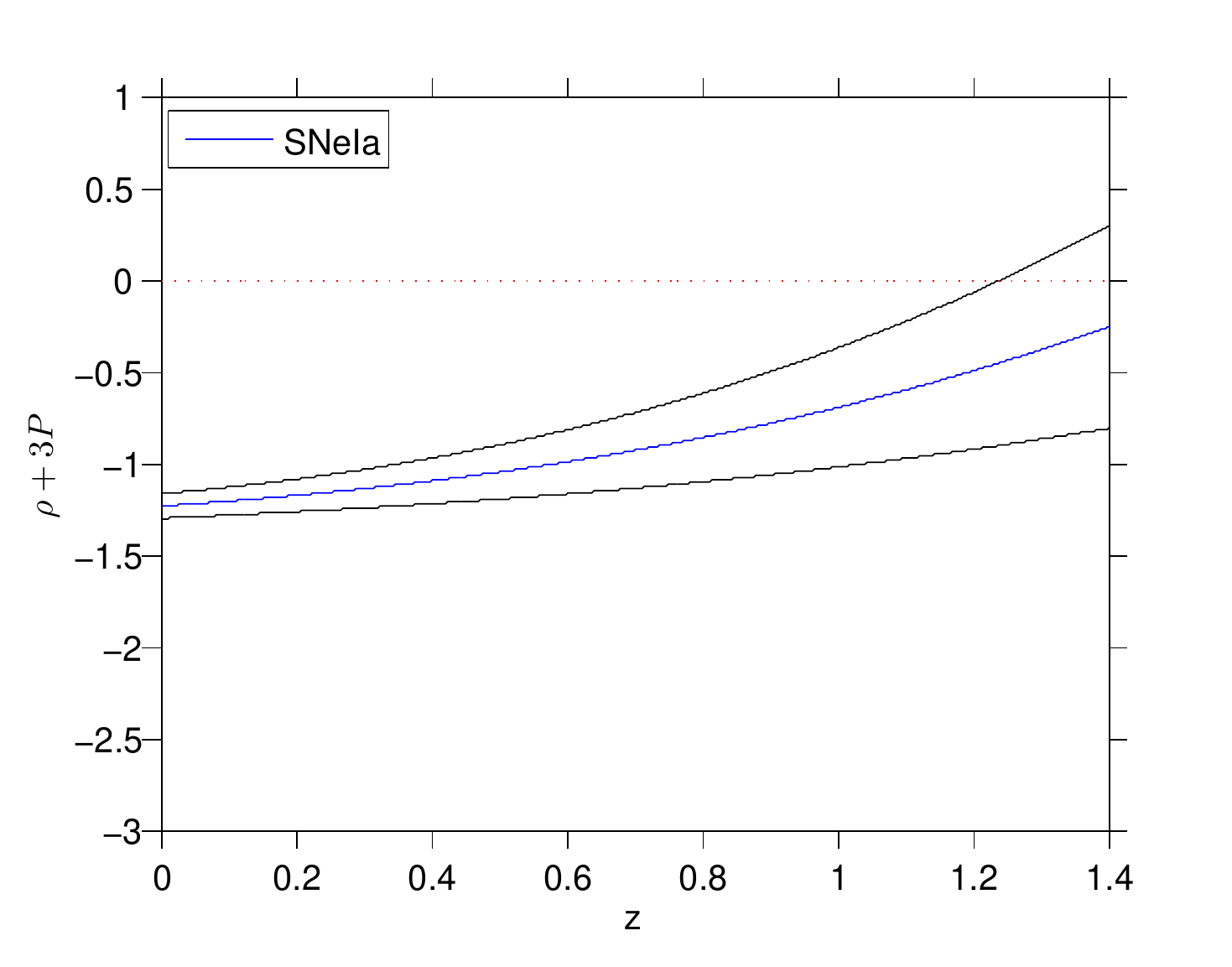}\label{fig61a}} \subfloat[Part
3][] {\includegraphics[width=3in]{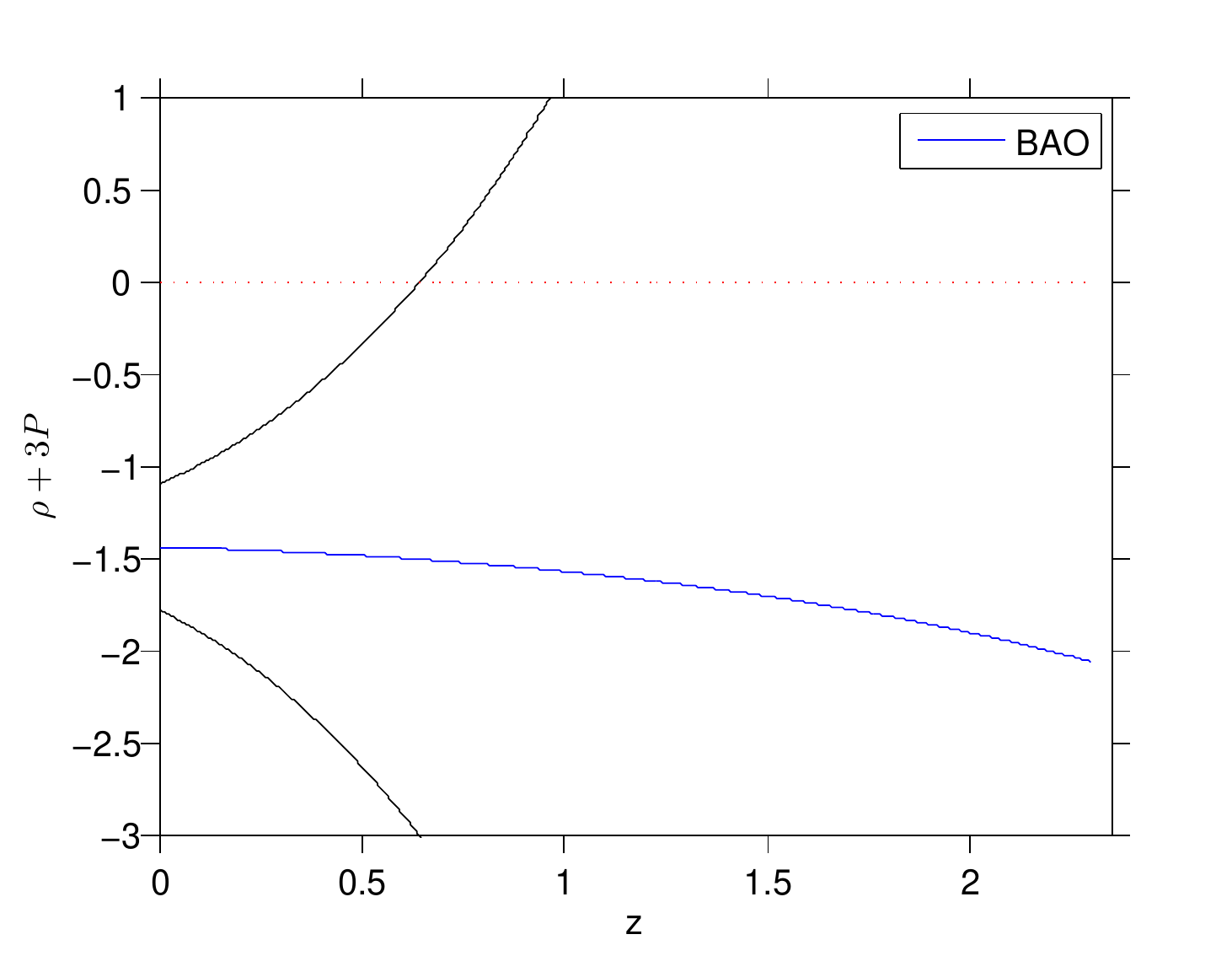}\label{fig62a}} 
\caption{Solid lines show the reconstructed $\rho+3P$ as a function of redshift with 1$\sigma$ error bars (black curves). In figure (a) the dashed lines corresponds 2$\sigma$ error bars.}
\label{figfa}
\end{figure}
The estimates of the model parameters obtained from SNeIa and BAO data are given below.\\
 
\noindent{\bf SNeIa}: $\Omega_m$ = 0.275 $\pm$ 0.033, $\Delta_{\mu}$ =-0.022$\pm$ 0.033 and $\tau^2$ = 0.839$\pm$ 0.114. The posteriors for $\Omega_m$ and $\Delta_{\mu}$, are fairly well peaked. The posterior of $\tau^2$ peaks at 0.917. $\sigma_f$ and $l$ have non-Gaussian posterior and there is high degeneracy between the two. Almost all values in the prior range are allowed. As discussed before it is preferable to marginalize these parameters since optimization may not give robust results.\\

\noindent{\bf BAO}: $H_0$ = 68.789 $\pm$6.691 and $\Omega_m$ = 0.3$\pm$0.114. As before the posteriors for $\sigma_f$ and $l$ are non-Gaussian and there is high degeneracy between these parameters. 

\subsection{Dark energy dynamics}
Many different kinds of potentials have been described in the
literature to explain the evolution of the scalar field \cite{sami}. 
To understand the characteristics of these potentials which can describe the accelerated expansion, is a big challenge. Direct reconstruction of the potential from the observational data is a better approach to understand the physics behind the accelerated expansion of the Universe. There have been many attempts in this direction. Luminosity distances of high redshift supernovae have been used in a parametric setting to constrain the dark energy potential \cite{saini,depot}. Parametrized form of effective equation of state of dark energy, deceleration parameter and dark energy density has been used to reconstruct quintessence potential \cite{guo}. Huterer and Peiris described the scalar field potentials by using a polynomial series and using PCA obtained constraints on the  parameters of the potential \cite{pei}. We summarise the results of our analysis of the DE variables below.
\begin{enumerate}[(a)]
\item We show the reconstruction of the potential $V(z)$, and $\dot{\phi^2}$, in Figs. (\ref{figaa}) and (\ref{figba}), respectively (in units of critical density).
\begin{figure}[ht]
\centering \subfloat[Part 3][]{
\includegraphics[width=3in]{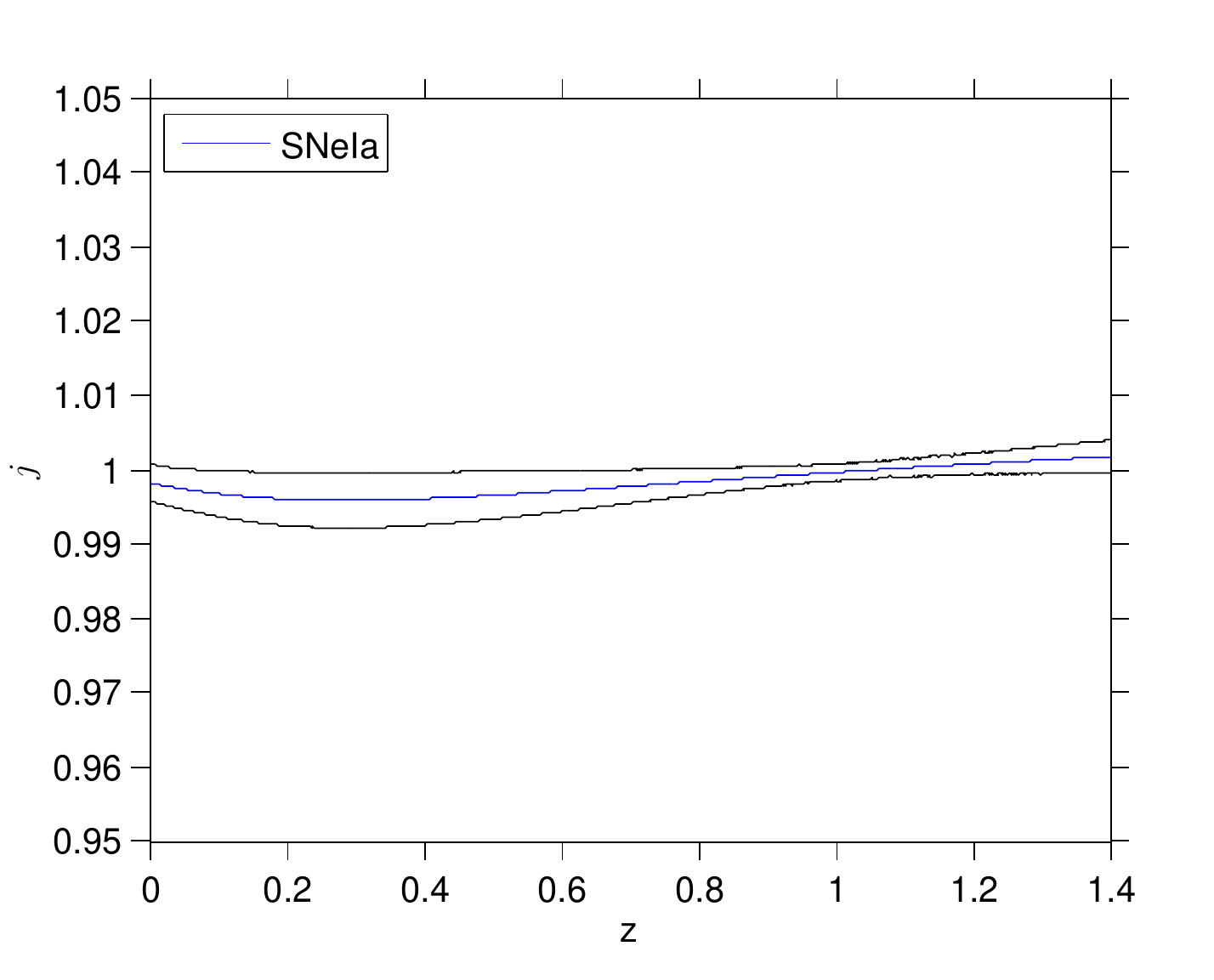}\label{fig81a}} \subfloat[Part
3][]
{\includegraphics[width=3in]{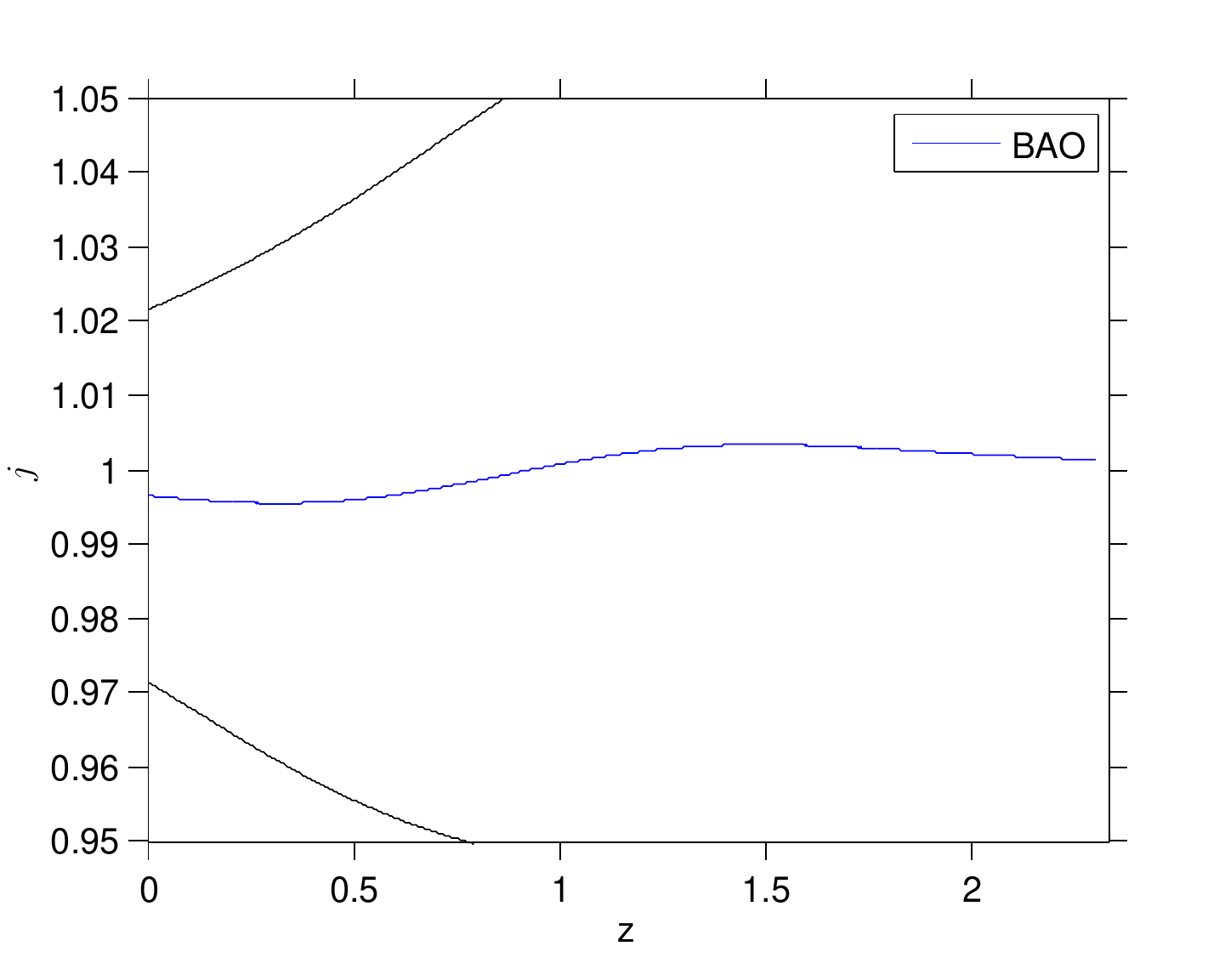}\label{fig82a}} 
\caption{Solid lines show the reconstructed jerk parameter as a function of redshift with 1$\sigma$ error bars (black curves).}  
\label{figha}
\end{figure}
As expected SNeIa data gives better constraints, specially at low redshift ($z < 0.5$) compared to  BAO data, since the number of data points is much larger. 
\item In case of variables which depend explicitly on $\Omega_m$ ($V$, $\dot{\phi}^2$, $\rho+3P$ and $\omega$), the error budget is effected by the estimated variance of $\Omega_m$, which is much larger in BAO data. 
\item
In slow roll approximation the kinetic energy of the scalar field approaches zero. This feature is recovered from the SNeIa data at 2$\sigma$ and BAO data at 1$\sigma$ (Fig. \ref{figba}). 
\item
The reconstructed  eos $\omega$ (Fig \ref{figea}) is consistent with the cosmological constant model of dark  energy ($\omega =-1$), at 2$\sigma$ level for SNeIa data and at 1$\sigma$ for BAO data. Our result agrees with the reconstruction of eos by \cite{hols} within error bars and we observe their method (of using MCMC within GP modelling to directly reconstruct $\omega$) gives tighter constraint on $\omega$. 
\item
We also reconstruct the SEC using the SNeIa and BAO 
observations (Fig \ref{figfa}). We find that it is violated ($\rho+3P<0$) for almost the entire redshift range of SNeIa at 1$\sigma$. For BAO data, $\rho+3P>0$ is allowed at 1$\sigma$ for a large redshift range but we observe that the error bars are too large to derive a conclusion. The NEC is equivalent to $\dot{\phi}^2\geq0$, and we observe that it is satisfied in the entire redshift range of both data sets (Fig \ref{figba}) within 2$\sigma$ for SNeIa data and within 1$\sigma$ for BAO data. Here too the error bars from BAO are very large.
\end{enumerate}
\subsection{Kinematic variables: $j$ and $\epsilon_H$}
\begin{figure}[ht]
\centering \subfloat[Part 3][]{
\includegraphics[width=3in]{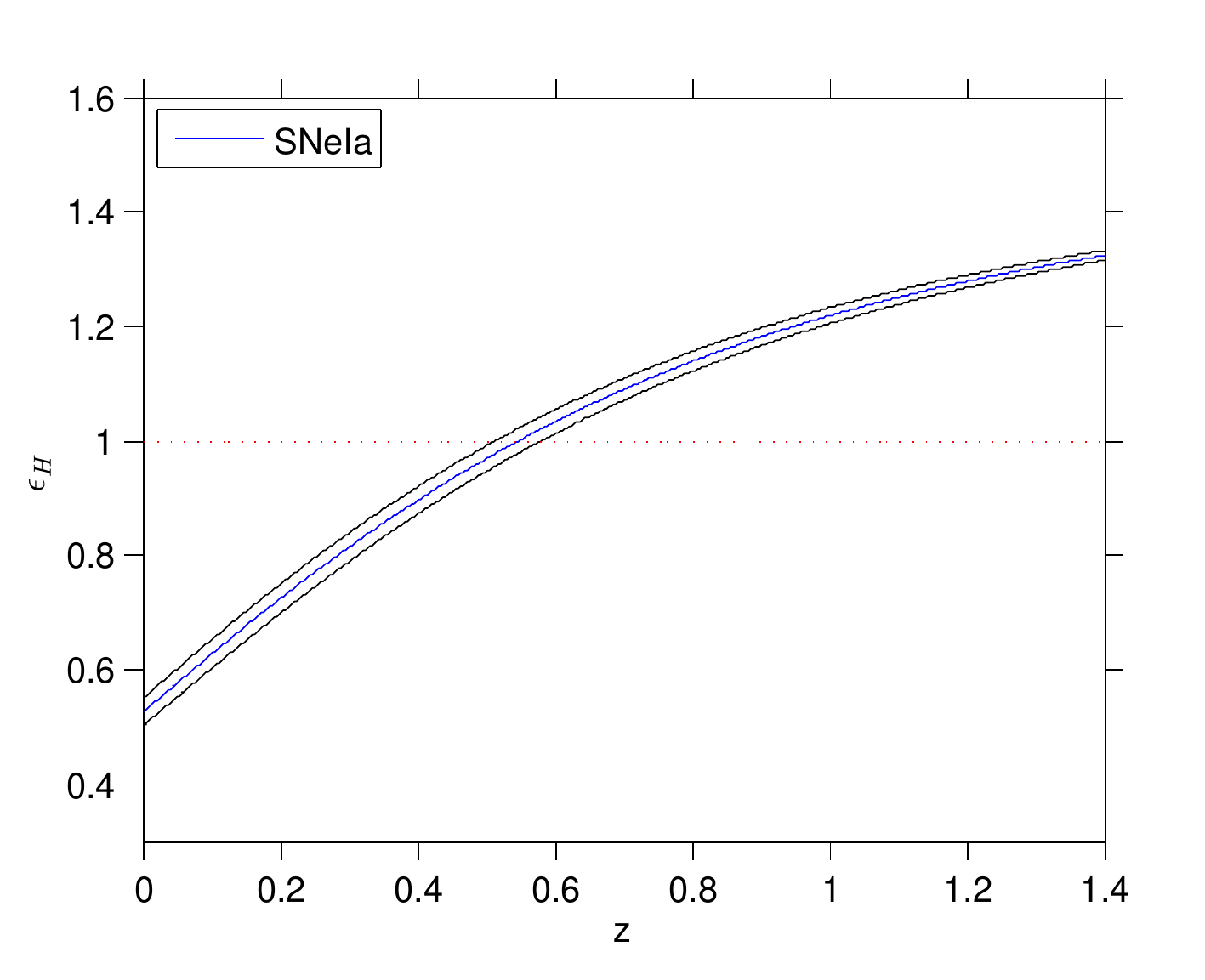}\label{fig71a}} \subfloat[Part
3][] {\includegraphics[width=3in]{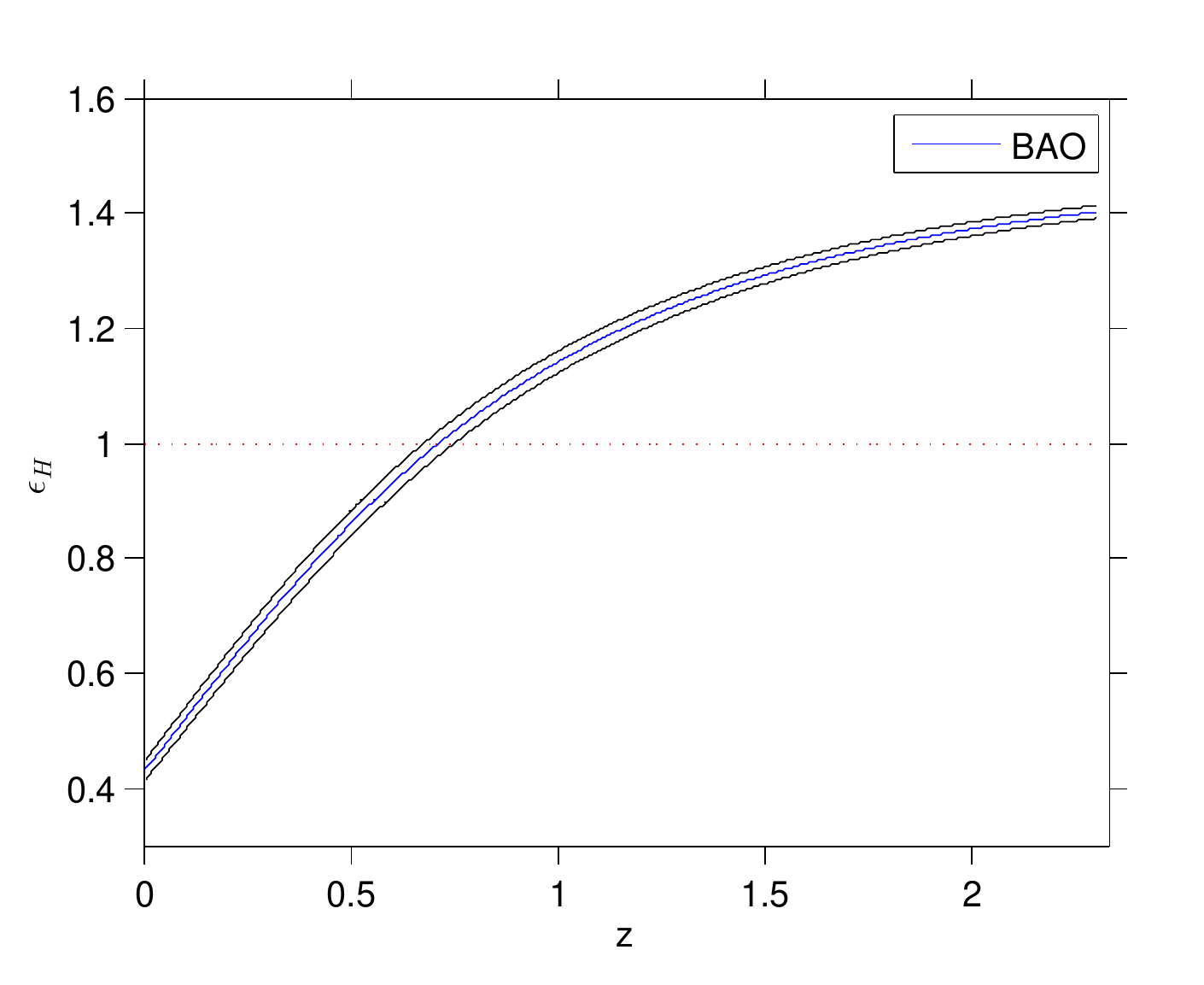}\label{fig72a}} 
\caption{Solid lines show the reconstructed $\epsilon_H$ as a function of redshift with 1$\sigma$ error bars (black curves).} 
\label{figga}
\end{figure}

Here we observe that the reconstructed error bars on both the kinematic variables are very tight, especially in case of SNeIa data. Here it is worth noting that the majority of the data points is in the near redshift range. So one would expect the error bars to get larger as $z$ increases. This is not evident from the plots of the kinematic variables. As mentioned earlier, the error bars on these variables are obtained through error propagation, where the errors on the reconstructed functions, are weighted by the partial derivatives of the variables, with respect to the functions reconstructed ($\partial f/\partial H$ etc.). These weights are redshift dependent and this may be reflected in the overall evolution of the errors on the dynamical variables.
\begin{enumerate} [(a)]
\item
For $\Lambda$CDM model $j = 1$, and deviation from this value would indicate a possible departure from $\Lambda$CDM. Barotropic models (where pressure 
of DE depends strictly on the energy density, i.e. $P = f(\rho)$) like 
generalized Chaplygin gas, satisfy $j \geq 1$ while for quintessence 
models $j \leq 1$ (\cite{chap}). The reconstructed jerk parameter from both data sets agrees well with 
the $\Lambda$CDM model (Fig. \ref{figha}). 
\item
The other kinematic variable $\epsilon_H$, is related to the deceleration parameter ($\epsilon_H$ = 1 + $q$). Hence, the coasting 
point where $\epsilon_H$ = 1 (i.e. $q$ = 0), gives the transition redshift.  SNeIa data and BAO give 
$z_t \approx 0.54$ and $z_t \approx 0.71$ respectively (Fig. \ref{figga}). Here we note that \cite{shaf} reconstructed the deceleration parameter $q$, using GP and found constraints using mock as well as real (SNeIa) data. Now, since $\epsilon_H$ and $q$ are related by a constant, one can compare these two reconstructions. We observe that, our error bars are tighter than those reported by \cite{shaf}. In their analysis Arman et al., have accounted for the multiple scale nature of the reconstruction by opting for a smoothing procedure to derive the initial mean function. They use five initial guesses that cross different redshift ranges and cover a wide range of behaviour. In this regard, their error bars may be more robust than ours.
\end{enumerate}

\acknowledgments Authors thank the anonymous referee whose valuable suggestions have greatly improved the paper. It is a pleasure to thank Philipp Hennig and David Duvenaud for insightful discussions. One of the author (D.J.) thanks A. Mukherjee and S. Mahajan for providing the facilities to carry out the research. R.N. acknowledges support under CSIR - SRF scheme (Govt. of India). SJ acknowledges support under UGC minor research project (42-1068/2013(SR)).


\begin{thebibliography}{30}

\bibitem{de} Frieman J. et al., {\it Dark Energy and the Accelerating Universe}, 
~\ARAnA ~{\bf 46} (2008) 385 [arXiv:0803.0982];\\
Caldwell R.R. \& Kamionkowski M., {\it The Physics of Cosmic 
Acceleration}, {\it Ann. Rev. Nucl. Part. Sci.} ~{\bf 59} (2009) 397
[arXiv:0903.0866].

\bibitem{sah} Sahni V.\& Starobinsky A.,
{\it Reconstructing Dark Energy}, ~\ijmpd ~{\bf 15} (2006) 2105
[arXiv:astro-ph/0610026].

\bibitem{sami} Copeland E.J. et al., {\it Dynamics of dark energy},
~\ijmpd ~{\bf 15} (2006) 1753 [hep-th/0603057].

\bibitem{sne} Supernova Search Team collaboration, Riess A.G. et al., 
{\it Observational evidence from supernovae for an accelerating universe 
and a cosmological constant}, ~\aj ~{\bf 116} (1998) 1009 
[astro-ph/9805201];\\
Supernova Cosmology Project collaboration, Perlmutter S. et al., {\it 
Measurements of Omega and Lambda from 42 high redshift supernovae}, 
~\apj ~{\bf 517} (1999) 565 [astro-ph/9812133];\\
SNLS collaboration, Astier P. et al., {\it The Supernova legacy survey: 
Measurement of $\Omega _m$, $\Omega _\Lambda$ and $w$ from the first 
year data set}, ~\AnA ~{\bf 447} (2006) 31 [astro-ph/0510447].

\bibitem{accn} Weinberg D.H. et al., {\it Observational Probes of 
Cosmic Acceleration} [arXiv:1201.2434] 

\bibitem{surv} Sloan Digital Sky Survey (SDSS): http://www.sdss3.org;\\
Giant Magellan Telescope (GMT): http://www.gmto.org;\\
James Webb Space Telescope (JWST): www.jwst.nasa.gov;\\
Euclid survey: http://sci.esa.int/euclid;\\
Dark Energy Survey: http://sci.esa.int/euclid;\\
Large Synoptic Survey Telescope (LSST): http://www.lsst.org;\\

\bibitem{wein} Weinberg S., {\it The Cosmological Constant Problem}, 
{\it Rev. Mod. Phys.} ~{\bf 61}, (1989) 1.

\bibitem{kunz} Kunz M., {\it The phenomenological approach to modeling 
the dark energy}, {\it Comptes Rendus Physique} ~{\bf 13} (2012) 539 
[arXiv:1204.5482].

\bibitem{bas} Bassett B. A. et al., {\it The essence of quintessence 
and the cost of compression}, ~\apjl ~{\bf 617} (2004) L1 
[astro-ph/0407364].

\bibitem{vinc} Vitagliano V. et al., {\it High-redshift cosmography}, 
~{\it JCAP}~{\bf 1003} (2010) 005 [arXiv:0911.1249];\\
Xia J-Q. et al., {\it Cosmography beyond standard candles and rulers}, 
~\prd ~{\bf 85} (2012) 043520 [arXiv:1103.0378].

\bibitem{kine} Cai R.G. et al., {\it Probing the dynamical behavior 
of dark energy}, {\it JCAP} ~{\bf 04} (2010) 012 [arXiv:1001.2207];\\
Lu J. et al., {\it Constraints on kinematic models from the latest 
observational data}, ~\plb ~{\bf 699} (2011) 246 [arXiv:1105.1871];\\
Xu L. \& Wang Y., {\it Cosmography: Supernovae Union2, Baryon Acoustic 
Oscillation, Observational Hubble Data and Gamma Ray Bursts}, ~\plb 
~{\bf 702} (2011) 114 [arXiv:1009.0963];\\
Nair R. et al., {\it Cosmokinetics: A joint analysis of Standard Candles, 
Rulers and Cosmic Clocks}, {\it JCAP} ~{\bf 01} (2012) 018 
[arXiv:1109.4574];\\
Wang S. et al., {\it Exploring the Latest Union2 SNIa Dataset by Using
Model-Independent Parametrization Methods}, ~\prd ~{\bf 83} (2011)
023010 [arXiv:1009.5837];\\
Giostri R. et al., {\it From cosmic deceleration to acceleration: 
new constraints from SN Ia and BAO/CMB}, {\it JCAP} ~{\bf 03} (2012) 
027 [arXiv:1203.3213];\\
Campo S. del et al., {\it Three thermodynamically-based 
parameterizations of the deceleration parameter}, ~\prd ~{\bf 86} 
(2012) 083509 [arXiv:1209.3415];\\
Sendra I. \& Lazkoz R., {\it SN and BAO constraints on (new) 
polynomial dark energy parametrizations: current results and forecasts}, 
\mnras ~{\bf 422} (2012) 776 [arXiv:1105.4943];\\
Neben A.R. \& Turner M.S., {\it Beyond $H_0$ and $q_0$: Cosmology is 
no longer just two numbers} [arXiv:1209.0480].

\bibitem{sahni} 
Sahni V. et al., {\it Statefinder-A new geometrical diagnostic of dark energy}, ~{\it JETP} ~{\bf 77} (2003) 207;\\
Alam U. et al., {\it Exploring the expanding Universe and dark energy using 
the statefinder diagnostic}, ~\mnras ~{\bf 344} (2003) 1057 
[astro-ph/0303009];\\
Arabsalmani M \& Sahni V. {\it Stateﬁnder hierarchy: An extended null diagnostic for concordance cosmology}, ~\prd ~{\bf 83} (2011) 043501.

\bibitem{om}
Sahni V. et al., {\it Two new diagnostics of dark energy}, 
~\prd ~{\bf 78} (2008) 103502.

\bibitem{om3sh} Shafieloo A et al., {\it New null diagnostic customized for reconstructing 
the properties of dark energy from baryon acoustic oscillations data},
~\prd ~{\bf 86} (2012) 103527.

\bibitem{ywang} Wang Y. \& Tegmark M., {\it Uncorrelated measurements 
of the cosmic expansion history and dark energy from supernovae}, 
~\prd ~{\bf 71} (2005) 103513 [astro-ph/0501351].

\bibitem{trot} Zunckel C. \& Trotta R., {\it Reconstructing the 
history of dark energy using maximum entropy}, ~\mnras ~{\bf 380} 
(2007) 865 [astro-ph/0702695].

\bibitem{hut} Huterer D. \& Starkman G., {\it Parameterization of 
dark-energy properties: A Principal-component approach}, ~\prl 
~{\bf 90} (2003) 031301 [astro-ph/0207517].

\bibitem{shap} Shapiro C. \& Turner M.S., {\it What do we really know 
about cosmic acceleration?}, ~\apj ~{\bf 649} (2006) 563 
[astro-ph/0512586].

\bibitem {jason} Dick J. et al., {\it Reduction of cosmological data 
for the detection of time-varying dark energy density}, 
{\it JCAP} ~{\bf 07} (2006) 001 [astro-ph/0603247];\\
Nair R. \& Jhingan S., {\it Is dark energy evolving?}, {\it JCAP} 
~{\bf 02} (2013) 049 [arXiv:1212.6644].

\bibitem{amara} Kitching T. D. \& Amara A., {\it Fisher matrix decomposition for dark energy prediction}, ~\mnras ~{\bf 398} (2009) 2134. 

\bibitem{put} Putter R.D. \& Linder E., {\it Being PC: Principal 
Components and Dark Energy}, [arXiv:0812.1794].

\bibitem{arman1} 
Shafieloo A. et al., {\it Smoothing Supernova Data to Reconstruct the Expansion History of the Universe and its Age}, ~\mnras ~{\bf 366} 
(2006) 1081 [astro-ph/0505329];\\
Shafieloo A., {\it Model Independent Reconstruction of the Expansion 
History of the Universe and the Properties of Dark Energy}, 
~\mnras ~{\bf 380} (2007) 1573 [astro-ph/0703034];\\
Shafieloo A. \& Clarkson C., {\it Model independent tests of the 
standard cosmological model}, ~\prd ~{\bf 81} (2010) 083537 
[arXiv:0911.4858].

\bibitem{arman2}
Shafieloo A. et al., {\it The Crossing Statistic: 
Dealing with Unknown Errors in the Dispersion of Type Ia Supernovae}, 
~{\it JCAP} ~{\bf 08} (2011) 017 [arXiv:1006.2141];\\
Shafieloo A., {\it Crossing Statistic: Bayesian interpretation, model selection and resolving dark energy parametrization problem}, 
~{\it JCAP} ~{\bf 05} (2012) 024 [arXiv:1202.4808].

\bibitem{arman}
Shafieloo A., {\it Crossing statistic: reconstructing the
expansion history of the universe}, ~{\it JCAP} ~{\bf 08} (2012)
002 [arXiv:1204.1109].

\bibitem{nes} Nesseris S \& García-Bellido J., {\it A new perspective on 
dark energy modeling via genetic algorithms}, ~{\it JCAP} ~{\bf 11}
(2012) 033 [arXiv:1205.0364].

\bibitem{hols} Holsclaw T et al., {\it Nonparametric Reconstruction 
of the Dark Energy Equation of State}, ~\prd ~{\bf 82} (2010) 103502 
[arXiv:1009.5443].\\
Holsclaw T et al., {\it Nonparametric Dark Energy Reconstruction 
from Supernova Data},~\prl ~{\bf 105} (2010) 241302 
[arXiv:1011.3079].\\
Holsclaw T et al., {\it Nonparametric Reconstruction of the Dark 
Energy Equation of State from Diverse Data Sets}, ~\prd ~{\bf 84} 
(2011) 083501 [arXiv:1104.2041].

\bibitem{seikel} Seikel M. et al., {\it Reconstruction of dark energy 
and expansion dynamics using Gaussian processes}, {\it JCAP} ~{\bf 06} 
(2012) 036 [arXiv:1204.2832]. 

\bibitem{shaf} Shafieloo et al., {\it Gaussian process cosmography}, 
~\prd ~{\bf 85} (2012) 123530 [arXiv:1204.2272].

\bibitem{fable} Crittenden R. G. et al., {\it Fables of reconstruction:
controlling bias in the dark energy equation of state}, {\it JCAP} 
~{\bf 02} (2012) 048 [arXiv:1112.1693].

\bibitem{saini} Saini T.D. et al., {\it Reconstructing the cosmic 
equation of state from supernova distances}, ~ \prl ~{\bf 85} (2000) 
6 [astro-ph/9910231]. 

\bibitem{join} Li C. et al., {\it Direct reconstruction of the dark 
energy scalar-field potential}, ~ \prd ~{\bf 75} (2007) 103503 
[astro-ph/0611093],\\ 
Sahlen M., {\it Direct reconstruction of the 
quintessence potential}, ~\prd ~{\bf 72} (2005) 083511 
[astro-ph/0506696],\\
Simon J. et al., {\it Constraints on the redshift 
dependence of the dark energy potential}, \prd ~{\bf 71} (2005) 
123001 [astro-ph/0412269],\\
Martinez E.F. \& Verde L., {\it Prospects for 
constraining the dark energy potential}, {\it JCAP} ~{\bf 08} (2008) 023 [arXiv:0806.1871],\\
Vazquez J.A. et al., {\it Reconstruction of the dark 
energy equation of state}, {\it JCAP} ~{\bf 09} (2012) 020 
[arXiv:1205.0847].

\bibitem{planck} Planck Collaboration, {\it Planck 2013 results. XVI. 
Cosmological parameters}, [arXiv:1303.5076].

\bibitem{HE} S.W. Hawking and G. F. R. Ellis, \emph{The Large Scale
Structure of Space-Time} (Cambridge University Press,
Cambridge, England, 1973).

\bibitem{wald} R. Wald, \emph{General Relativity} (University of Chicago,
Chicago, IL, 1984).

\bibitem{caroll} S. Carroll, \emph{Spacetime and Geometry} (Addison-Wesley,
Pearson, 2004).


\bibitem{matt} Visser M., {\it General Relativistic Energy Conditions: 
The Hubble expansion in the epoch of galaxy formation}, ~\prd ~{\bf 56} (1997) 7578 [gr-qc/9705070].

\bibitem{ec} Santos J. et al., {\it Energy conditions and cosmic acceleration} 
~\prd ~{\bf 75} (2007) 083523 [astro-ph/0702728];\\
Lima M.P. et al., {\it Energy condition bounds and their confrontation 
with supernovae data}, ~\prd ~{\bf 77} (2008) 083518;\\
Santos J. et al., {\it Energy conditions and supernovae observations}
~\prd ~{\bf 74} (2006) 067301;\\
Schuecker P. et al., {\it Observational constraints on general relativistic 
energy conditions, cosmic matter density and dark energy from X-ray clusters 
of galaxies and type-Ia supernovae}, ~\AnA ~{\bf 402} (2003) 53;\\
Gong Y. et al., {\it Energy conditions and current acceleration of the universe}, 
~\plb ~{\bf 652} (2007) 63;\\
Wu C-J. et al., {\it Reconstructing the history of energy condition violation 
from observational data}, ~\apj ~{\bf 753} (2012) 2.


\bibitem{rap} Rapetti D. et al., {\it A kinematical approach to dark 
energy studies}, ~\mnras ~{\bf 375} (2007) 1510 [astro-ph/0605683].

\bibitem{lima} Guimaraes A. C. C. et al., {\it Bayesian Analysis and Constraints 
on Kinematic Models from Union SNIa}, ~{\it JCAP} ~{\bf 10} (2009) 010 [0904.3550]. 

\bibitem{zhai} Zhai Z-X. et al., {\it Reconstruction and constraint of 
the jerk parameter from OHD and SNe Ia observations}, [arXiv:1303.1620].

\bibitem{rasm} Rasmussen C. and Williams C., {\it Gaussian Processes 
for Machine Learning}, MIT Press, Cambridge U.S.A. (2006); \\
Williams C., {\it Prediction with Gaussian processes: From linear 
regression to linear prediction and beyond, in Learning in Graphical 
Models}, M.I. Jordan eds., MIT Press, Cambridge U.S.A. (1999);\\
MacKay D., {\it Information Theory, Inference and Learning Algorithms}, 
Cambridge University Press, Cambridge U.K. (2003).


\bibitem{gazt} Gaztanaga E. et al., {\it Clustering of luminous red galaxies
 – IV. Baryon acoustic peak in the line-of-sight direction and a direct 
 measurement of H(z)}, ~\mnras ~{\bf 399} (2009) 1663 [arXiv:0807.3551].

\bibitem{wang} Chuang, C. H., \& Wang, Y., {\it Measurements of $H(z)$ 
and $D_A(z)$ from the two-dimensional two-point correlation function of 
Sloan Digital Sky Survey luminous red galaxies}, ~\mnras ~{\bf 426} (2012) 
226 [arXiv:1102.2251]. 

\bibitem{blake} Blake C. et al., {\it The WiggleZ Dark Energy Survey: Joint
measurements of the expansion and growth history at $z < 1$}, ~\mnras 
~{\bf 425} (2012) 405 [arXiv:1204.3674].

\bibitem{kazin} Kazin E. A. et al., {\it The Clustering of Galaxies in the 
SDSS-III Baryon Oscillation Spectroscopic Survey: Measuring $H(z)$ and 
$D_A(z)$ at $z = 0.57$ with Clustering Wedges} [arXiv:1303.4391].

\bibitem{busca} Busca N.G. et al., {\it Baryon acoustic oscillations in the
Ly$\alpha$ forest of BOSS quasars}, ~\AnA ~{\bf 552} (2013) A96 
[arXiv:1211.2616].

\bibitem{suz} Suzuki N. et al., {\it The Hubble Space Telescope Cluster 
Supernova Survey: V. Improving the Dark Energy Constraints Above $z > 1$ and 
Building an Early-Type-Hosted Supernova Sample}, ~\apj  ~{\bf 85} (2012)
[arXiv:1105.3470].

\bibitem{depot} Chiba T. \& Nakamura T., {\it Feasibility of reconstructing 
the quintessential potential using type Ia supernova data}, ~\prd
~{\bf 62} (2000) 121301;\\
Huterer D. \& Turner M.S., {\it Prospects for probing the dark energy via 
supernova distance measurements}, ~\prd ~{\bf 60} (1999) 081301.

\bibitem{guo} Wang Y. et al., {\it Reconstructing Dark Energy Potentials From
Parameterized Deceleration Parameters}, ~{\it Chin. Phys. B.} ~{\bf 19}
(2010) 019801 [arXiv:1004.3370];\\
Guo Z-K. et al., {\it Parametrization of quintessence and its potential},
~\prd ~{\bf 72} (2005) 023504;\\
Guo Z-K. et al., {\it Parametrizations of the Dark Energy Density and Scalar 
Potentials}, ~{\it Mod. Phys. Lett. A} ~{\bf 22} (2007) 883 
[astro-ph/0603109].

\bibitem{pei} Huterer D. \& Peiris H., {\it Dynamical behavior of generic 
quintessence potentials: Constraints on key dark energy observables}
~\prd ~{\bf 75} (2007) 083503.


\bibitem{chap} Chiba T. \& Nakamura T., {\it The Luminosity Distance, the 
Equation of State, and the Geometry of the Universe}, 
~{\it Prog. Theor. Phys.} ~{\bf 100} (1998) 1077 [astro-ph/9808022];\\
Linder E. V \& Scherrer R. J., {\it Aetherizing Lambda: Barotropic fluids 
as dark energy}, ~\prd ~{\bf 80} (2009) 023008 [arXiv:0811.2797];\\
Caldwell R.R \& Kamionkowski M., {\it Expansion, geometry, and gravity},
~{\it JCAP} ~{\bf 09} (2004) 009 [astro-ph/0403003];\\
Blandford R.D. et al., {\it Cosmokinetics}, ~{\it ASP Conference Series}
~{\bf 339} (2005) 27 [astro-ph/0408279].


\end{thebibliography}
\end{document}